\documentclass[prb,twocolumn,groupedaddress,superscriptaddress,showpacs,floatfix,longbibliography]{revtex4-2}

\usepackage{amsfonts}
\usepackage{graphicx}
\usepackage{dcolumn}
\usepackage{bm}
\usepackage{amsmath}
\usepackage{amssymb}
\usepackage{revsymb4-1}
\usepackage{hhline}
\usepackage[normalem]{ulem}

\usepackage{color}

\usepackage{hyperref}
\hypersetup{colorlinks=true,citecolor={blue},linkcolor={blue},urlcolor={blue}}

\begin{document}

\title{Exploring Structural Nonlinearity in Binary Polariton-Based Neuromorphic Architectures}

\author{Evgeny~Sedov}
\email{evgeny\_sedov@mail.ru}
\affiliation{Spin Optics Laboratory, St. Petersburg State University, Ulyanovskaya 1, St. Petersburg 198504, Russia}
\affiliation{Stoletov Vladimir State University, 87 Gorky str., Vladimir 600000, Russia}

\author{Alexey~Kavokin}
\affiliation{Moscow Institute of Physics and Technology, Institutskiy per., 9, Dolgoprudnyi, Moscow Region, 141701, Russia}
\affiliation{Spin Optics Laboratory, St. Petersburg State University, Ulyanovskaya 1, St. Petersburg 198504, Russia}
\affiliation{
School of Science, Westlake University, 600 Dunyu Road, Hangzhou 310030, Zhejiang Province, China}

\begin{abstract}
This study investigates the performance of a binarized neuromorphic network leveraging polariton dyads, optically excited pairs of interfering polariton condensates  within a microcavity to function as binary logic gate neurons.
Employing numerical simulations, we explore various neuron configurations, both linear (NAND, NOR) and nonlinear (XNOR), to assess their effectiveness in image classification tasks. 
We demonstrate that structural nonlinearity, derived from the network's layout, plays a crucial role in facilitating complex computational tasks, effectively reducing the reliance on the inherent nonlinearity of individual neurons.
Our findings suggest that the network's configuration and the interaction among its elements can emulate the benefits of nonlinearity, thus potentially simplifying the design and manufacturing of neuromorphic systems and enhancing their scalability.
This shift in focus from individual neuron properties to network architecture could lead to significant advancements in the efficiency and applicability of neuromorphic computing.
\end{abstract}

\maketitle

\section{Introduction}

Artificial neural networks (ANNs) have revolutionized data processing by emulating the intricate network of neurons in the human brain, enabling significant advances in fields ranging from robotics to healthcare~\cite{Ozmen2021,healthcare12020125}.
These systems process information through interconnected nodes or neurons that can learn to perform complex tasks, leading to improvements in decision-making and pattern recognition technologies.
As the demand for these technologies grows, so does the interest in developing various hardware implementations to support them~\cite{APRes7011312}.
These hardware platforms include electronic-based systems, which leverage silicon-based technologies, and photonic systems, which exploit the interaction of light and matter to enhance speed and reduce energy consumption~\cite{Science345668,Nanolett203506,PhysRevApplied18024028, SciRep77430,PhysRevApplied21064027}.
Another promising option is the use of exciton-polaritons, quasiparticles that combine the properties of light and matter.
These are being investigated for their potential in neuromorphic computing, particularly because of their rapid operation times and potentially low power consumption~\cite{NanoLett213715,PhysRevApplied16024045}.
Each of these platforms aims to offer unique advantages, whether in scalability, speed, or energy efficiency, to meet the growing computational demands of modern ANNs.

In light of the 2024 Nobel Prize in Physics awarded to John Hopfield and Geoffrey Hinton for foundational advances that have shaped the modern era of neural networks and machine learning, our exploration into polariton-based neuromorphic architectures gains added relevance.
Hopfield’s seminal contributions to the physics of exciton-polaritons~\cite{PhysRev1121555,PhysRev182945} and neural network theory~\cite{Hopfield1982aa,science3755256} have inspired new approaches that blend these fields.
This fusion of knowledge is at the heart of our study, underscoring the potential of exciton-polaritons in neuromorphic computing to push the boundaries of processing speeds and the inherent capability for parallel data handling.

Binarized neural networks (BNNs) represent a specific approach to enhancing the computational efficiency of artificial neural networks~\cite{electronics8060661,SciRep911705,QIN2020107281}.
By simplifying the weights and activations within the network to just two levels, typically 0 and 1, BNNs drastically reduce the computational complexity and memory usage required for neural processing.
Although this simplification often results in lower accuracy compared to networks with full-precision weights, BNNs excel in scenarios where speed, power efficiency, and low resource consumption are more critical than achieving the highest possible accuracy, making them well-suited for applications in internet of things, edge computing, and other environments where autonomy and limited resources are key considerations~\cite{QIN2020107281,Chi2021Logic,Liang2018FPBNN}.

Binarized neural networks have been effectively realized using exciton-polaritons.
In the notable implementation described in Ref.~\cite{NanoLett213715}, artificial neurons function as XOR gates.
The used technique utilizes nonresonant laser pulses, acting as the input signals, to selectively excite spatially localized exciton-polariton condensates, that interact with each other.
The resulting output signals vary in energy, reflecting the different combinations of the inputs.
This approach has proven successful in pattern recognition tasks, achieving approximately 96\% accuracy on the MNIST (Mixed National Institute of Standards and Technology) dataset, a standard benchmark in machine learning for handwritten digit recognition, under noisy conditions using a single-hidden-layer network. 
The impressive potential of this solution is further underscored by subsequent assessments of its remarkable energy efficiency, as reported in~\cite{PhysRevApplied16024045}.

Nonlinearity is a cornerstone in the operational efficiency of neural networks, essential for executing tasks beyond the scope of linear computational models.
This includes distinguishing overlapping data sets or solving inherently complex problems.
The nonlinear activation function within each neuron exemplifies this intrinsic nonlinearity, defining how inputs are transformed into outputs in a way that linear operations cannot~\cite{DUBEY202292}.
Exciton polaritons, known for their pronounced nonlinear properties due to polariton-polariton interactions, are especially valuable in this context.
The distinctive nonlinearity of polaritons is the key element that drives the functionality of both continuous-weight networks~\cite{Nanolett203506} and binarized neuromorphic systems~\cite{NanoLett213715}.

Recent research~\cite{NatPhysWanjura2024, PNAS120e2305027120,xia2023deeplearningpassiveoptical, yildirim2024nonlinearprocessinglinearoptics} challenge the emphasis traditionally placed on the inherent nonlinearity in individual neurons within neural networks, see also~\cite{NatPhysMcMahon2024,NatRevElEng1358}.
Studies have demonstrated that nonlinear computations can be realized using purely linear optical systems by adjusting the parameters of these systems.
This development underscores that achieving nonlinearity does not necessarily rely on the physical nonlinearity of the system's components.
By encoding inputs as parameters rather than direct signals, a linear system can emulate nonlinear behavior.
This approach shifts the focus from the inherent properties of the materials to the configuration of the system itself, which facilitates \textit{structural nonlinearity} arising from the arrangement and interactions among its components.

In our recent paper~\cite{sedov2024polarlattneur}, we have theoretically proposed a binarized neuromorphic network architecture based on a lattice of pairwise coupled exciton polariton condensates.
In this geometry, each pair of condensates, referred to as a polariton dyad~\cite{PhysRevLett124207402}, serves as artificial binary neurons functioning similarly to OR gates.
Unlike XOR gate neurons utilized in work~\cite{NanoLett213715}, the OR operation is linear.
Nevertheless, in~\cite{sedov2024polarlattneur}, we demonstrated that our proposed architecture effectively addresses the inherently nonlinear challenge of image classification, exemplified by the recognition tasks in the MNIST dataset.
Our current  study elucidates the role of structural nonlinearity in solving recognition tasks.
We explore the potential for modifying the operation of polariton neurons proposed in~\cite{sedov2024polarlattneur} to function as both linear (NAND and NOR) and nonlinear (XNOR) gates.
Through numerical experiments, we compare the image classification accuracies, allowing us to question whether the significance of inherent nonlinearity, typical of individual computational elements such as neurons, might be overstated.

\section{Polariton dyads as artificial neurons}

\subsection{Introduction to polariton dyads}

\begin{figure*}[tb!]
\begin{center}
\includegraphics[width=0.8\linewidth]{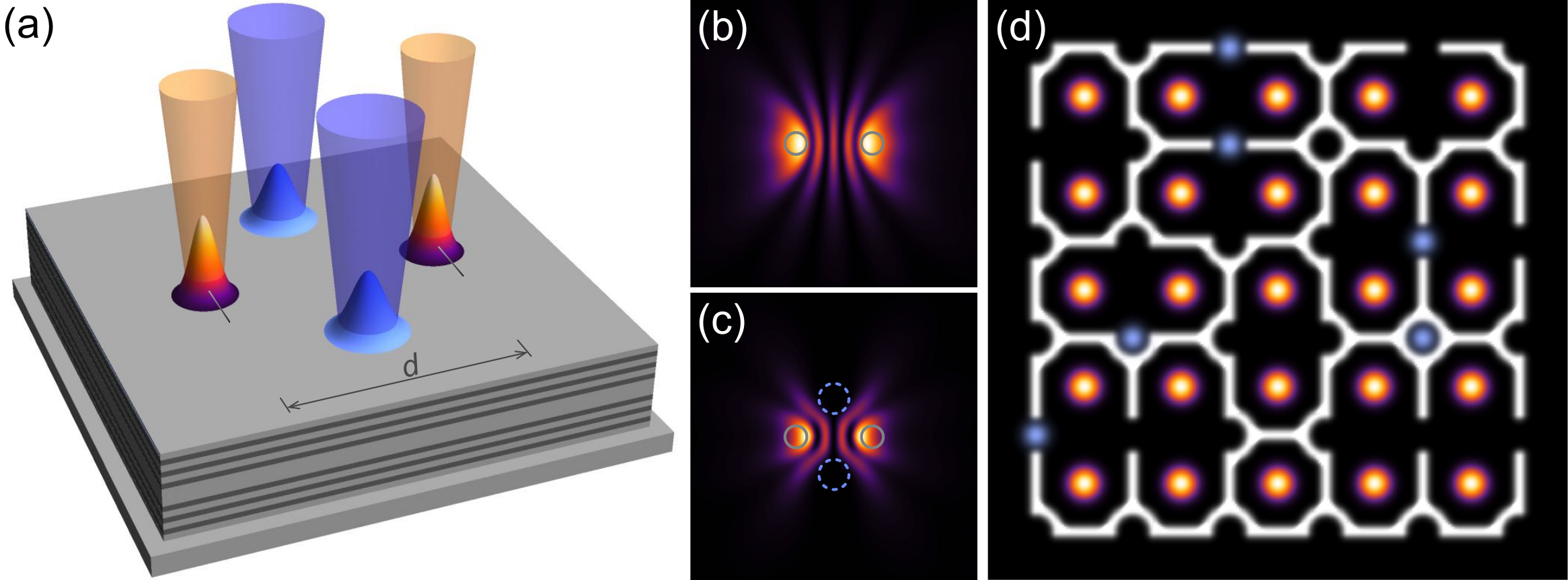}
\end{center}
\caption{
(a) Schematic of the excitation and control of a polariton dyad within an optical microcavity, using localized non-resonant laser pumping.
Orange cones represent the laser pump for excitation of the condensates, while blue cones indicate the control signals.
(b,c) Schematic density distribution of polaritons within the dyad in the absence (b) and presence (c) of the control signals.
The positions of the excitation and control laser beams are indicated by gray solid and blue dashed circles, respectively.
(d) Example of the effective potential geometry, along with pump and control signal beams for exciting a lattice of polariton dyads.
}
\label{FIG_TaskScheme}
\end{figure*}

In the present work, we theoretically explore a polariton neuromorphic network, an idea initially proposed in our recent study~\cite{sedov2024polarlattneur}.
The concept centers around a regular spatial lattice of pairwise coupled exciton-polariton condensates --- polariton dyads --- forming the structural backbone of our system.
Each dyad is formed by two polariton condensates within a planar semiconductor microcavity, excited through spatially localized, non-resonant laser beams separated by a distance~$d$, see schematic of excitation of a polariton dyad in~Fig.~\ref{FIG_TaskScheme}(a).
The excitation of these condensates is carried out in a non-resonant regime, implying that the energy of the laser pump is significantly higher than the energy of polariton eigenmodes of the microcavity.
This pump facilitates the creation of an incoherent exciton reservoir at higher energy, from which polaritons are subsequently stimulated to scatter into lower energy condensate states within the microcavity, see detailed explanation of non-resonant pumping in polariton systems, e. g., in~\cite{Nature443409,ACSPhot71163}.

Coherence between the condensates is established through the exchange of ballistic polaritons, which are characterized by large wave vectors.
These polaritons propagate with finite velocities across the plane of the microcavity, effectively mediating interactions between condensates.
The spatial localization of the excitation source and the repulsive interaction with the exciton reservoir under the pump spots create a potential landscape that facilitates the downhill flow of polaritons, enhancing their mobility and interaction potential.
As a result of this dynamic, an interference pattern emerges between the coherently linked condensates.
Given the proximity of the condensates within each dyad, the interference fringes are pronounced and readily detectable, see Fig.~\ref{FIG_TaskScheme}(b).
The coherence length in a system of coupled condensates significantly exceeds the dimensions of individual condensates, as discussed in~\cite{PhysRevLett124207402,Optica8106}.

The interference pattern within each polariton dyad is critically influenced by the phase relationships between the condensates, which dictate whether the interference is constructive or destructive.
Both conservative and non-conservative processes contribute to the formation of interference patterns.
To describe these processes, an effective complex potential can be introduced.
In practice, the control of interference patterns is achieved by optically manipulating the effective potential landscape through non-resonant excitation of spatially localized incoherent exciton reservoirs within the dyad area.
This manipulation allowing one to toggle between even and odd interference patterns (with maximum and minimum density at the midpoint between the condensates).

In our earlier work~\cite{sedov2024polarlattneur}, we followed a methodology that targeted the non-conservative component of the effective potential, as proposed in~\cite{PhysRevLett124207402}.
We suggested using pump beams with circular polarization orthogonal to that of the beams exciting the condensates, aiming to excite optically active (bright) exciton reservoirs.
Interaction of quasiparticles with orthogonal circular polarizations is negligible~\cite{PhysRevB72075317,PhysRevB90195307}, which results in a reduced contribution to the conservative component of the potential.
However, spin relaxation mechanisms still allow the orthogonally polarized reservoir to partially feed the condensate state~\cite{NJPhys20075008,PhysRevLett128117401,PhysRevLett124207402}, thus impacting the non-conservative component of the potential and affecting the interference patterns.

\begin{figure*}[tb!]
\begin{center}
\includegraphics[width=\linewidth]{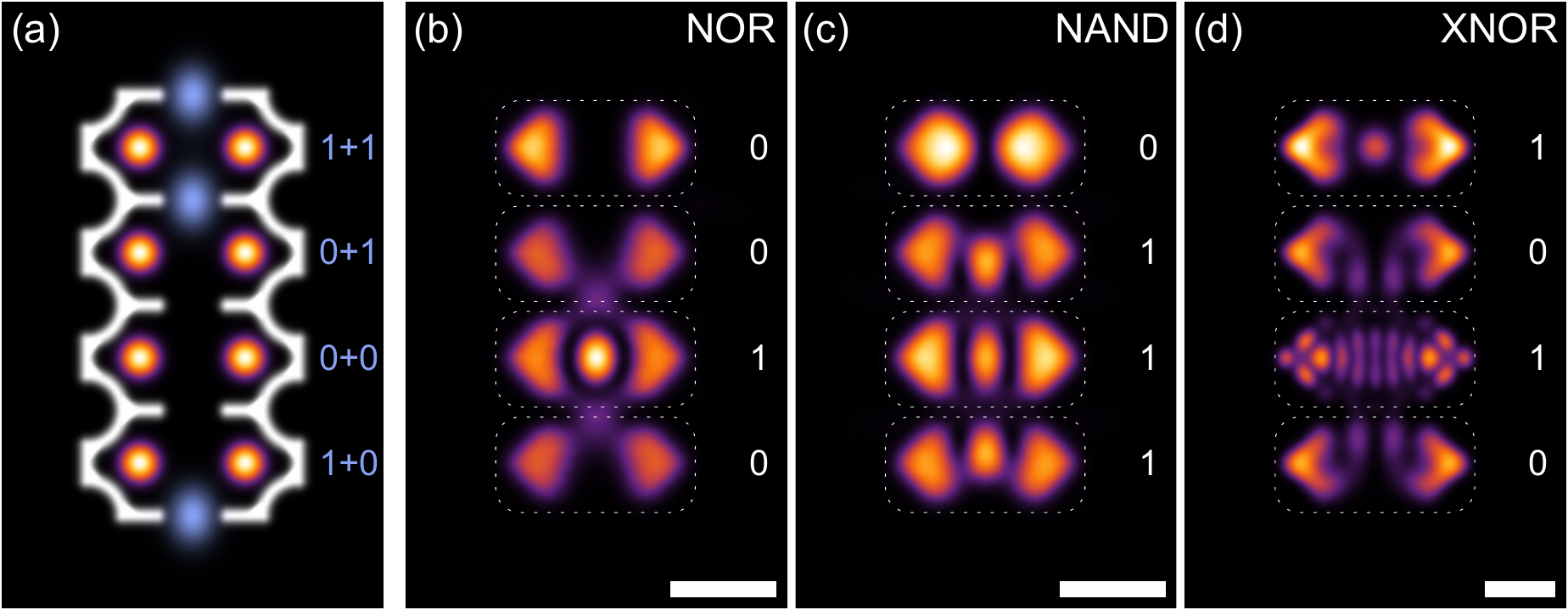}
\end{center}
\caption{
(a) Schematic representation of an assembly  of four binary polariton neurons displaying various input signal combinations, from top to bottom: `1 and 1', `0 and 1', `0 and 0' and `1 and 0'.
The primary color scheme illustrates the profiles of non-resonant optical pumping of condensates within the dyads.
The potential trap profile, isolating individual polaritonic dyads, is depicted in white.
Control laser beams, acting as input signals, are shown in blue.
Panels (b), (c), and (d) show spatial distribution of polaritons in the presence of control signals within neurons configured to function based on NOR, NAND, and XNOR gates, respectively.
White dashed boxes serve as guides for the eye to outline the area of individual polariton neurons.
The output value for each neuron, reflects whether there is a signal present in the neuron's center (1) or not (0).
A white bar at the bottom of the panels corresponds to a scale of~$10\, \mu\text{m}$.
}
\label{FIG_Gates}
\end{figure*}

Building on this foundation, we have opted in this study for a different approach by exciting optically inactive (dark) exciton reservoirs.
The optical excitation of dark excitons is discussed in Refs.~\cite{PhysRevLett122047403,PhysRevB108165411}.
Dark excitons, though not contributing directly to the polariton population due to their inability to couple to photons within the microcavity, provide a substantial influence through their repulsive interactions with the polaritons.
This interaction considerably alters the potential landscape, offering a precise mechanism to control the spatial coherence and thus the operational state of the dyads.

\subsection{Interference control and dyad isolation}

To control the interference patterns within each polariton dyad, we suggest to locally excite dark exciton reservoirs using a pump beam positioned equidistantly from the condensates.
This placement is strategic for our square lattice configuration, where each pump beam can affect two adjacent dyads simultaneously.
Additionally, each dyad can be influenced by two such beams, as illustrated in Fig.~\ref{FIG_TaskScheme}(a).
Adjusting the pump intensity distribution allows one to alter the parity of the interference pattern within the dyads, cf. Figs.~\ref{FIG_TaskScheme}(b) and~\ref{FIG_TaskScheme}(c).

In our model, the dyads serve as artificial neurons, with ON and OFF states corresponding to maximum and minimum intensity at the center between the condensates, respectively.
The pump beams exciting the reservoirs act as inputs.
However, they affect the neuron states rather than directly act as processing signals.
This approach uses input signals as control instruments that adjust the interaction dynamics within the dyads, akin to the principles of structural nonlinearity observed in linear photonic networks~\cite{NatPhysWanjura2024, PNAS120e2305027120,xia2023deeplearningpassiveoptical, yildirim2024nonlinearprocessinglinearoptics, NatPhysMcMahon2024,NatRevElEng1358} and outlined~by~us~in~\cite{sedov2024polarlattneur}.

To ensure the operational integrity of the polariton neuromorphic network, isolating each dyad from its neighbors is crucial.
Without isolation, unintended interactions could lead to phase locking, compromising the designed binary neuron functionality.
This interaction generally results from direct coupling between the condensates rather than through the controlled input signals.

Several strategies can be utilized to ensure isolation between dyads in a polariton neuromorphic network.
Creating effective potential barriers between dyads is crucial to prevent undesirable interactions.
These barriers can be established either by modifying the microcavity structure itself or by creating optical traps~\cite{PhysRevB92035305,PhysRevB97235303,PhysRevB107045302} using focused laser beams.
In analogy to the non-resonant excitation of signals, optical induction of dark exciton reservoirs can be employed to form potential barriers.
Further, the microcavity structure can be modified to physically separate the condensates.
Barriers can be established through techniques such as etching, which structures microcavities to create isolated areas~\cite{PhysRevLett108126403,PhysRevLett116066402}.
Additional methods include varying the dissipation profiles and polariton lifetimes through techniques such as electrical carrier injection or stress application to the substrate, which dynamically adjust interaction dynamics within the dyad network~\cite{AdvQT31900065,RepProgrPhys80016503}.
All these approaches ensure effective isolation, with the choice of method largely dictated by the specifics of the experimental setup and technological capabilities.

Figure~\ref{FIG_TaskScheme}(d) provides an example of a potential barrier configuration for a lattice of dyads, illustrating the positioning of pump spots for dyad excitation (main color scheme), signal inputs (blue), and the layout of potential barriers (white) to ensure necessary isolation.
In this configuration, the polariton dyads are arranged randomly, forming the hidden layer of the polariton neural network.
This random arrangement fosters a variety of inter-neuron relationships, mediated by input signals.
Namely, a single neuron may be influenced by one or two signals, or none at all, while each signal may impact one or two neurons, or none whatsoever.
This combination of flexibility and randomness lays the foundation for structural nonlinearity within the network.
As will be demonstrated, among the multitude of possible random configurations, a specific arrangement that optimizes the functionality of the polariton network can be selected.
Nonetheless, all random configurations maintain structural nonlinearity, effectively surpassing the performance of purely linear network architectures.

\subsection{Logic gates as binary artificial neurons}

As discussed previously, polariton dyads can exhibit distinct states, OFF and ON, based on the parity of their interference patterns.
This capability allows the dyad to function as an OR gate, as explored through numerical simulations in our study~\cite{sedov2024polarlattneur}.
Such behavior can be viewed in terms of binary logic states: the dyad remains in the OFF state, showing a minimum intensity at the center of the dyad, in the absence of control signals.
Conversely, the presence of at least one control signal triggers a transition to the ON state, characterized by a maximum intensity.
Beyond the basic OR-gate functionality, the intricate interactions within coherent polariton condensates suggest the potential for a broader spectrum of logical operations.

Let us consider the geometry of the model problem as schematically illustrated in Fig.~\ref{FIG_Gates}(a).
Here, four isolated polariton dyads are arrayed in sequence, each separated by a consistently shaped potential barrier (depicted in white).
This barrier's consistent shape across each dyad facilitates the creation of various random configurations within the condensate lattice.
Three input signals (depicted in blue) are strategically placed to enable comprehensive coverage of all possible signal combinations affecting the binary gates.
Specifically, the top dyad receives two signals, simulating a dual `1' input condition.
The second top and the bottommost dyads each respond to a single signal, corresponding to `0 and 1' and `1 and 0' inputs, respectively.
The third dyad is not influenced by any signal, representing a `0 and 0' scenario.
This configuration effectively demonstrates how a polariton dyad can function under different input conditions, underscoring their potential as binary logic elements.

In this study, we consider negative (or inverted) logic gates that inherently remain in an ON state, signifying a `1' with maximum intensity at the center, when no input signals are present.
This configuration is primarily due to the previously discussed specific influence of control signals on the parity of interference patterns within dyads.
Namely, here we specifically address the effects of a repulsive optically induced real barrier on polaritons, which makes a negligible contribution to non-conservative processes, in contrast to the approach discussed in our previous work~\cite{sedov2024polarlattneur}.

To model the operation of an ensemble of polariton dyads, we solve the generalized Gross-Pitaevskii equation for the polariton wave function $\Psi (t,\mathbf{r})$, see details of the model in the Appendix.
We track the spatial distribution of polariton density within each dyad under their specific excitation conditions.
Figure~\ref{FIG_Gates}(b) illustrates the operation of a polariton dyad configured to function as a NOR gate.
The parameters that can be adjusted include the distance $d$ between the condensates in a dyads (lattice period), the geometry of the cross-section, and the intensity of the pump beams for both the condensates and the input pulses. 
The simulation results reveal that in the absence of input signals, the dyad (third from the top) remains in the ON state, characterized by a pronounced intensity maximum at its center.
Conversely, the presence of at least one signal switches the dyad to the OFF state, showing a minimum intensity in the region of interest.
A different configuration of control parameters enables the implementation of a NAND gate using a polariton dyad, as depicted in Fig.~\ref{FIG_Gates}(c).
In this configuration, the dyad switches its state only when both input signals are present, which is evident when comparing the top dyad to the others in the panel.

One can also see that spatial patterns of the interacting condensates in the trap differ from those without the trap, cf. individual dyads in Fig.~\ref{FIG_Gates}(b,c) with Fig.~\ref{FIG_TaskScheme}(b,c).
The potential modifies the spatial distribution of polaritons within the condensates, introducing effects of spatial quantization.
This particular phenomenon can be exploited to construct a XNOR gate based on a polariton dyad, as shown in Fig.~\ref{FIG_Gates}(d).
In the chosen configuration, while the corresponding dyad (second from the bottom) is in an even state with a central maximum in the absence of input signals, its state notably differs from those in Fig.~\ref{FIG_Gates}(b) and~\ref{FIG_Gates}(c).
Specifically, it corresponds to a higher quantization level of the confining potential along the dyad's major axis, characterized by an increased number of nodes.
The presence of a single input signal shifts the dyad to an odd state (OFF).
However, when both input signals are present, the dyad returns to an even state, albeit with fewer nodes than in the initial state.
While this difference is significant for the evolution of the polariton system, it remains inconsequential for the gate's functional integrity.

Thus, polariton dyads can be effectively tuned to function as  specific binary logic gates.
By ensuring equal shape of potential traps across the dyads, one can facilitate their combination in any arbitrary configuration within the lattice.
This allows for the flexible design of a network of dyad-based gates, enabling the architecture to be tailored to specific processing requirements.

\begin{figure*}[tb!]
\begin{center}
\includegraphics[width=\linewidth]{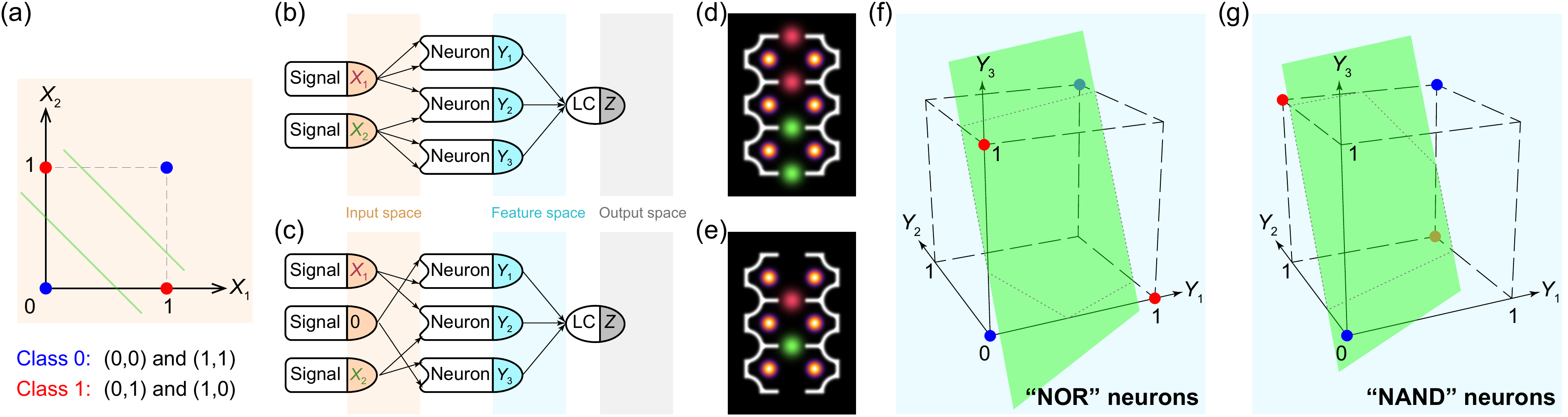}
\end{center}
\caption{
(a) The XOR problem in the input space: the two classes of outputs, 0 (blue) and 1 (red), cannot be separated by a single straight line. 
(b,c) Schematics of two simple neural networks with a hidden layer composed of three neurons, considered for solving the XOR problem.
(d,e) Schematics of the hidden layer with configurations of input signals as per (b) and (c), respectively.
The input signals $X_1$ and $X_2$ are denoted in maroon and green, respectively. 
(f,g) The XOR problem in the feature space of the hidden layers composed of three neurons functioning as NOR and NAND gates, respectively.
The solution in~(f) is effective for both configurations as in~(b) and~(c), while~(g) works solely for the configuration~in~(b).
The green plane separates the classes 0~and~1.
Dashed lines serve as guides for the eye.
}
\label{FIG_XORLinears}
\end{figure*}

\section{Neural network with linear binary neurons for solving the XOR problem}

The presence of nonlinearity in a neuromorphic computational system can be assessed by its ability to solve problems that are inherently nonlinear, e.g., classifying overlapping data sets, extracting features from complex patterns, or performing tasks that require the separation of intertwined data points.
If a system is limited only to linear tasks, it is generally categorized as a linear classifier, which may not be sufficient for more complex decision-making needs.
A typical benchmark for evaluating a system's nonlinear capabilities is the XOR problem. 
This involves two binary inputs that produce a single binary output, true (1) only when the inputs differ. 
The outcomes for all possible input pairs are: $(0,0)$ and $(1,1)$ yield 0, while $(0, 1)$ and $(1,0)$ yields 1. 
The XOR problem challenges the capacity of linear classifiers due to its requirement for a non-linear decision boundary. 
Linear classifiers can only separate data points using a straight line, which is insufficient for XOR where no single line can correctly classify all input pairs, see schematic in Fig.~\ref{FIG_XORLinears}(a). 
This illustrates the necessity for non-linear processing to handle more complex logical operations effectively.

In the study~\cite{NanoLett213715}, the XOR problem is tackled by employing the nonlinear properties of polariton condensates, configuring each neuron within the network to function as a XOR gate.
This approach underscores the unique capabilities of polaritonic systems, where each neuron does not merely pass signals but actively participates in nonlinear computations.
In the present study, we are focusing on a different type of nonlinearity that stems from the structure of the neural network, rather than from individual neurons.
Figures~\ref{FIG_XORLinears}(b) and~\ref{FIG_XORLinears}(c) show two model neural networks that utilize linear binary neurons to solve the XOR problem.
Each model incorporates an input layer, a hidden layer with three linear binary neurons, and an output layer.
The network in Fig.~\ref{FIG_XORLinears}(b) processes two binary inputs, $X_1$ and $X_2$.
$X_1$ is directed towards both inputs of the first neuron and one input of the second neuron, while $X_2$ feeds the remaining input of the second neuron and both inputs of the third neuron.
The arrangement in Fig.~\ref{FIG_XORLinears}(c) introduces an additional dummy signal that replaces $X_1$ at one of the inputs of the first neuron and $X_2$ at one of the inputs of the third neuron.
The layout of the hidden layer adapted for our polaritonic system is shown in Figs.~\ref{FIG_XORLinears}(d,e).
The output signals from the hidden layer neurons, $Y_{1,2,3}$, undergo linear classification.
This architecture is designed to ensure that if properly configured, the network should clearly separate the input signals into their respective classes, 0 or 1, confirming the model's effective nonlinearity and its capacity to resolve the XOR problem.

The diagram in Fig.~\ref{FIG_XORLinears}(a) illustrates the XOR problem within the two-dimensional input space of $(X_1, X_2)$.
As data passes through the hidden layer, it is projected into a feature space, the dimensionality of which corresponds to the number of involved neurons, three in the considered case.
This transformation aims to simplify the differentiation of complex patterns by clarifying the relationships among data points.
The network's ability to resolve the XOR problem hinges on whether the classes, labeled as 0 and 1, can be linearly separated in this feature space.

\begin{figure*}[tb!]
\begin{center}
\includegraphics[width=\linewidth]{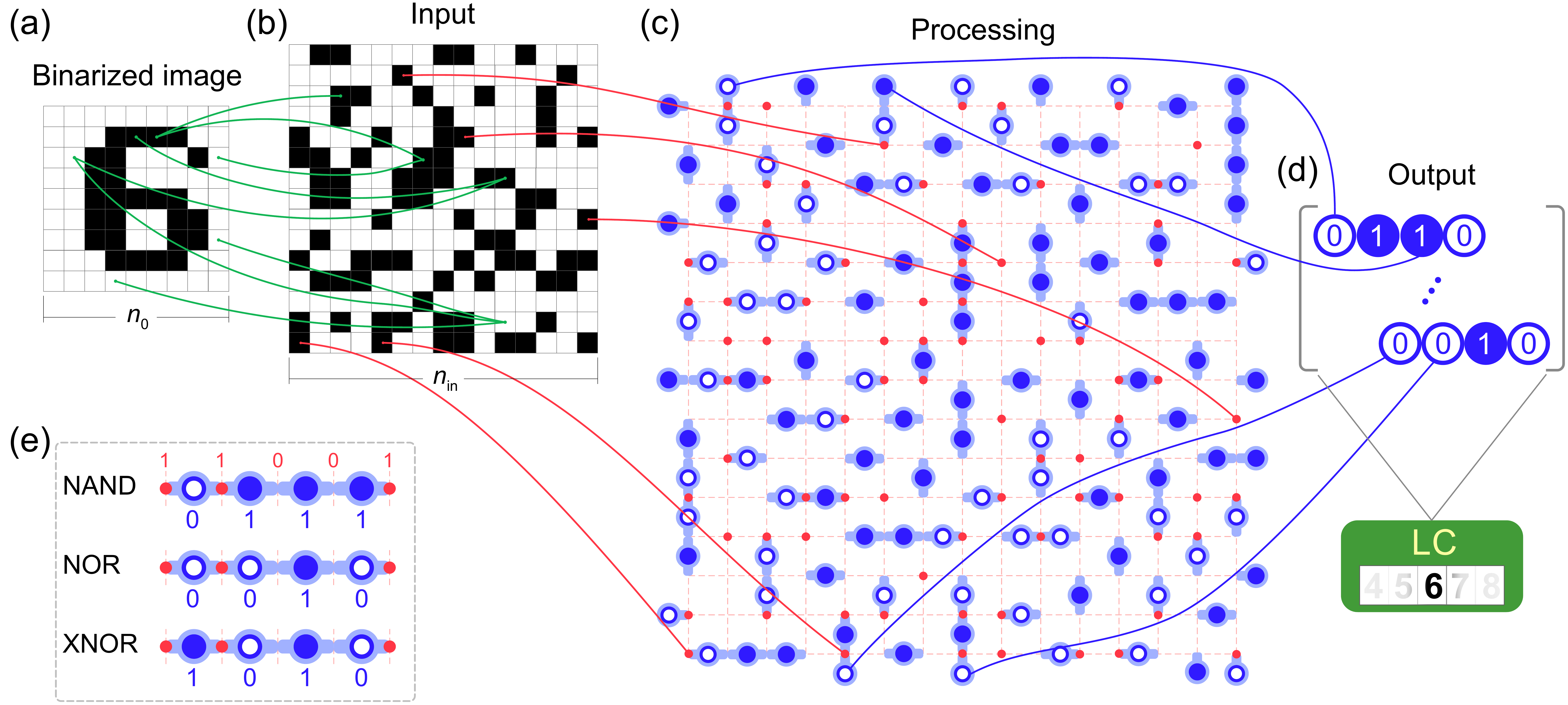}
\end{center}
\caption{
(a--d) Schematic depiction of a binary neural network configured with a lattice of pairwise coupled polariton condensates.
A binarized $28\times28$ pixel image from the MNIST dataset (a) is mapped onto a transformation lattice of dimensions $n_{\text{in}} \times n_{\text{in}}$ (b).
This lattice serves as the template for the input optical signal.
Neurons within the hidden layer are activated by this input~(c), generating the output signal.
This output is subsequently processed through a linear classifier LC~(d).
As an example, neurons in the diagram are shown functioning as XNOR gates.
(e) Schematic representation of the operation of artificial neuronal assemblies composed of NAND, NOR and XNOR gates is illustrated in Fig.~\ref{FIG_Gates}(b--d).
}
\label{FIG_Scheme}
\end{figure*}

Figures~\ref{FIG_XORLinears}(f) and~\ref{FIG_XORLinears}(g) display the complete set of possible input combinations projected into the feature space $(Y_1, Y_2, Y_3)$ at the output of the hidden layer, which comprises neurons functioning as OR gates in (f) and AND gates in (g).
For Fig.~\ref{FIG_XORLinears}(f), the projections are valid for both network architectures shown in Figs.~\ref{FIG_XORLinears}(b) and (c).
The scenario in Fig.~\ref{FIG_XORLinears}(g) applies only to the network configuration in Fig.~\ref{FIG_XORLinears}(b).
As illustrated, in both cases, the different classes of input data, labeled as 0 and 1 (denoted by blue and red dots respectively) can be successfully separated by a plane within this feature space and thus are linearly separable.
This outcome confirms that the XOR problem is successfully resolved using a network based on linear binary neurons, highlighting the fundamental capability of such networks to solve inherently nonlinear problems.

The next sections will explore the roles of structural and inherent neuron nonlinearity in solving such problems.
For complex tasks that require handling higher-dimensional feature spaces, straightforward visual interpretations become less feasible.
In our study, we will address this challenge by employing a benchmark problem, the classification of handwritten digits from the MNIST dataset, within a square lattice of binary neurons.
Our approach will include comprehensive numerical simulations aimed at evaluating and comparing the classification accuracy under different neuronal configurations.
Specifically, we aim to analyze how the network performs when neurons function as linear NOR or NAND gates, compared to their operation as nonlinear XOR gates.

\section{Network functionality and accuracy assessment}

Building on our understanding of the operation of individual polariton neurons and their activation principles within the hidden layer, we now explore the architecture of the neuromorphic network based on a polariton lattice.
To illustrate this architecture, we explore the task of classification of handwritten digits from the MNIST dataset, as depicted in Fig.~\ref{FIG_Scheme}.
The original image from the dataset is a grayscale bitmap with dimensions of $28 \times 28$ pixels, totaling 784 pixels.
This image is converted into a binary matrix of the same dimensions through a binarization process, where pixel values are rounded up, see Fig.~\ref{FIG_Scheme}(a).

In preparing the input signal pattern for the neuromorphic network, we implement two key operations concurrently: randomization and expansion, see Fig.~\ref{FIG_Scheme}(b).
During the randomization, elements from the binarized matrix are not allocated in a sequential manner; instead, they are placed into the new pattern matrix using a randomized approach.
This is achieved by applying a consistent randomization mask across all images.
Such a randomized distribution of input signals ensures that all neurons across the network are engaged more uniformly.
Simultaneously, the expansion process increases the size of the input matrix from $n_0 \times n_0$ to $n_{\text{in}} \times n_{\text{in}}$, enhancing the network's capability to process complex data.

The input matrix acts as the blueprint for generating the control signals in the hidden layer, see Fig.~\ref{FIG_Scheme}(c).
These signals represent optical pump pulses or beams, appropriately arranged using a spatial optical modulator or a precisely organized array of laser emitters.
These control signals modulate the potential landscape near each dyad, toggling them between ON and OFF states. 
Figure~\ref{FIG_Scheme}(c) exemplifies a network composed of neurons that operate as binary XNOR gates.
Schematic representations of NAND, NOR, and XNOR gates, responding to various combinations of control input signals, are depicted in~Fig.~\ref{FIG_Scheme}(e).

The state of the dyad is reflected by the absence or presence of photoluminescence at its center.
This optical response effectively encodes the binary output signal of the network.
The output signal is then processed through a conventional linear classifier, as shown in Fig.~\ref{FIG_Scheme}(d).

\begin{figure*}[tb!]
\begin{center}
\includegraphics[width=.5\linewidth]{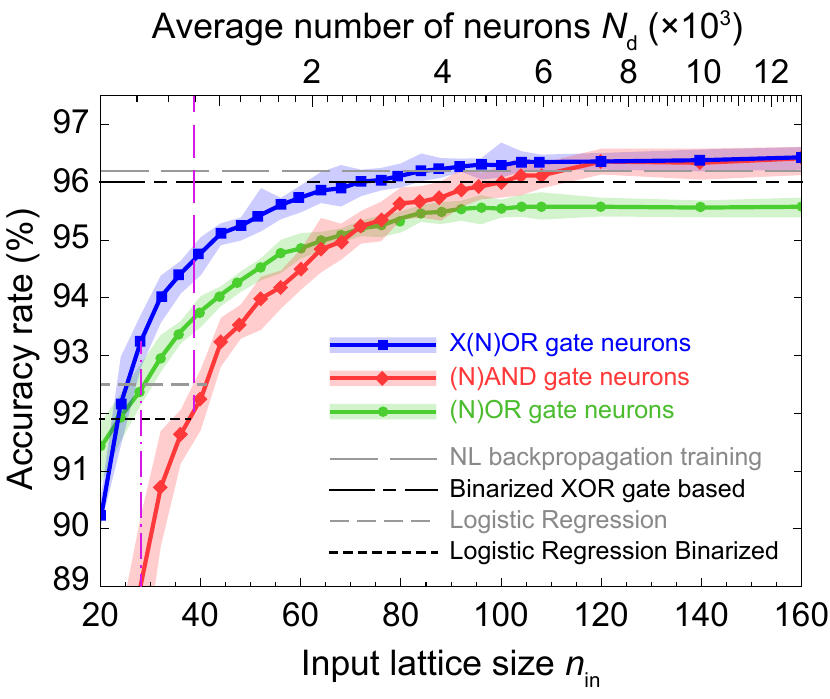}
\end{center}
\caption{
Dependence of the accuracy of MNIST handwritten digit recognition on the size of the input lattice $n_{\text{in}}$ (lower scale) or the number of neurons $N_{\text{d}}$ (upper scale), for neurons in the hidden layer functioning as XNOR (blue), NAND (red), and NOR (green) gates.
Each data point is the average outcome of ten numerical experiments, each employing distinct randomization masks.
The shaded area indicates the range of accuracy variation observed across these numerical experiments.
Vertical magenta lines mark specific conditions: dash-dotted line where the size of the polariton lattice corresponds to that of the initial images, and dashed line where the number of neurons matches the number of pixels in the initial image.
Horizontal dashed lines represent the accuracy levels of various established classification approaches: linear software classification for the grayscale (92.5\%) and the binarized (91.9\%) MNIST dataset, a binarized polariton network utilizing XOR gates~\cite{NanoLett213715}, and a nonlinear polariton network trained with software-based backpropagation~\cite{PhysRevApplied18024028}.
}
\label{FIG_Extens}
\end{figure*}

We simulate the operation of the discussed polariton neuromorphic network numerically to evaluate its functionality.
The standard assessment involves solving the task of handwritten digit recognition using the MNIST dataset as a benchmark.
This dataset, consisting of 60,000 training samples and 10,000 testing samples, serves as a widely recognized standard for image classification challenges, allowing us to rigorously compare our results with existing studies in polariton-based neuromorphic networks, as well as networks with various neuron configurations within our research.

Figure~\ref{FIG_Extens} illustrates the dependency of the accuracy of image classification on the size of the input signal lattice $n_{\text{in}}$ for neural networks composed of neurons functioning as NOR (green), NAND (red), and XNOR (blue) gates.
The upper horizontal scale in the figure represents an estimate of the number of neurons involved, correlated with the lattice size as ${N_{\text{d}} \le
\left(n_{\text{in}}^2 + 2 n_{\text{in}} + 1 \right)/2}$.
Notably, the green curve aligns with the dependence for the network with OR gates considered in~\cite{sedov2024polarlattneur}.
It is worth mentioning that similar correlations apply to other inverse gates, matching dependencies observed with direct gates.
This match occurs because, for the linear classifier, the inversion of signals 0 and 1 does not affect the outcome.

All dependencies exhibit a monotonically increasing tendency with the number of involved neurons and demonstrate a saturating character.
Notably, the curve for the neural network based on XNOR gate neurons rises most rapidly to its maximum value.
Remarkably, this maximum accuracy of 96\% aligns with that achieved by the binarized network based on XOR gates discussed in~\cite{NanoLett213715}.
The most noteworthy comparison from Fig.~\ref{FIG_Extens} reveals that while the NAND-based network lags at lower artificial neuron number, it ultimately achieves the same 96\% accuracy as the XNOR network at a comparable number of neurons.
This highlights the distinctive impact of neuron nonlinearity at lower counts, where the embedded nonlinearity of individual neurons plays a dominant role.
As the number of neurons increases, structural nonlinearity within the network becomes more pronounced, eventually overshadowing the contribution of individual neuron nonlinearity and becoming the dominant factor in network performance.

In Ref.~\cite{sedov2024polarlattneur}, a technique of input signal densing was introduced, aimed at enhancing the overall recognition accuracy of neural networks without altering the hidden layer's architecture.
Our forthcoming analysis will assess how this technique affects the efficiency of networks employing linear (NAND and NOR) versus nonlinear (XNOR) gate neurons.

The conventional approach presumes that each element in the input matrix directly corresponds to a single element in the binarized matrix.
The procedure of input signal densing modifies this approach by allowing multiple elements from the binarized matrix influence a single element of the input matrix.
This enhancement is achieved by aggregating the values of several elements to determine the value of one input element.
If any of the selected elements is a `0', the resulting element in the input matrix is set to~`0'.
This procedure emulates a logical NOR operation, which is particularly suitable for the inverse gate neurons discussed in this paper, similar to how the OR operation fits the direct gate neurons~\cite{sedov2024polarlattneur}.
This approach effectively increases the number of active input elements, thereby enhancing the network's ability to capture and process more complex patterns and interactions within the data.
To quantify the extent of this aggregation, we introduce the parameter densing degree~$s$, which defines the average number of binarized elements influencing each input matrix element.
For example, if $s=2$, it means that on average, each element in the input matrix is determined by two elements from the binarized matrix.

\begin{figure*}[tb!]
\begin{center}
\includegraphics[width=\linewidth]{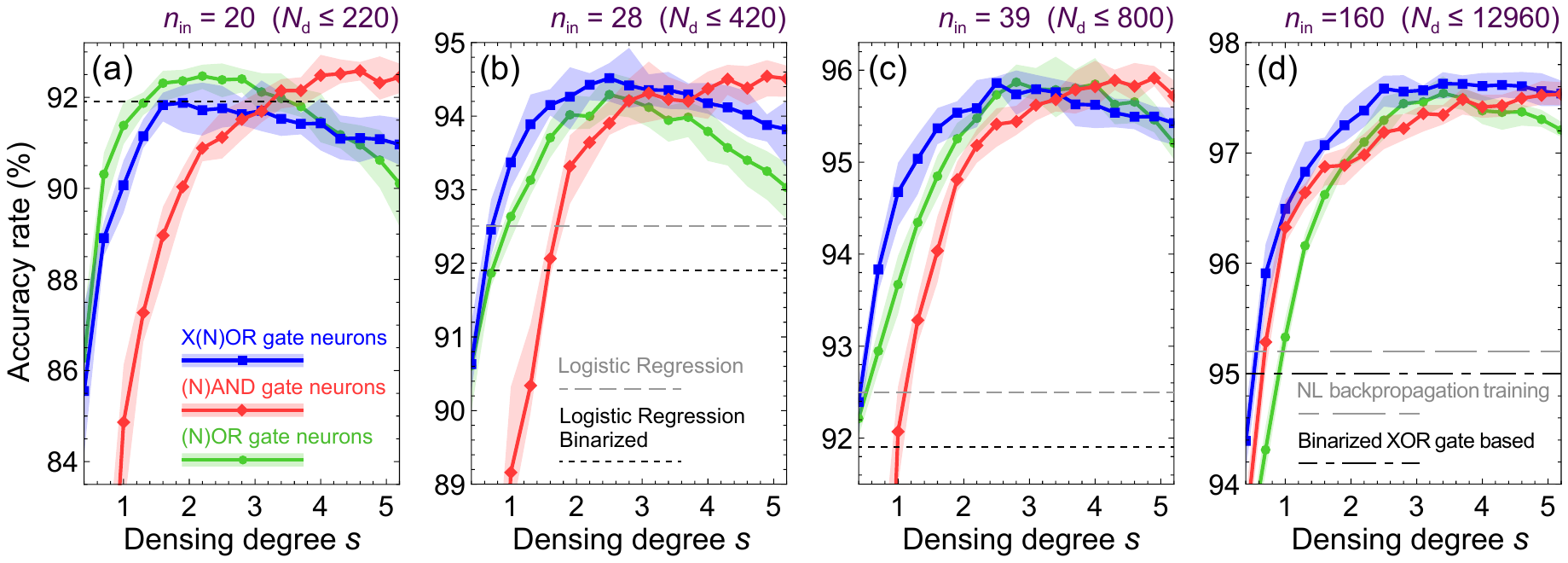}
\end{center}
\caption{
Dependence of the classification accuracy on the densing degree $s$ of the input signal for neurons in the hidden layer functioning as XNOR (blue), NAND (red), and NOR (green) gates for square polariton lattices of different size: $n_{\text{in}} = 20$ (a), $n_{\text{in}} = 28$ (b), $n_{\text{in}} = 39$ (c) and $n_{\text{in}} = 160$ (d).
Meaning of the horizontal dashed lines is the same as in Fig.~\ref{FIG_Extens}.
}
\label{FIG_Intens}
\end{figure*}

In Fig.~\ref{FIG_Intens}, we show the dependencies of recognition accuracy on the densing degree $s$ for neural networks utilizing neurons that function as various types of gates, both linear (NAND and NOR) and nonlinear (XNOR), across different sizes of the input matrix $n_{\text{in}}$. 
All dependencies observed across various lattice sizes $n_{\text{in}}$ characteristically feature pronounced peaks, indicating maximum accuracy for a given hidden layer architecture.
As the lattice size increases, the position of these peaks shifts towards higher values of~$s$.
The presence of these peaks and their dependence on $n_{\text{in}}$ are attributed to the balance between active and inactive neurons, which serve as the signal sources for subsequent processing by a linear classifier.
Optimal classification efficiency is expected when the numbers of active and inactive neurons are comparable.
Conversely, efficiency typically diminishes when one significantly outweighs the other, with comparable performance observed when the proportion of active to inactive neurons is reversed.

Across all figures, the rate of change in the accuracy on densing degree $s$ varies for networks composed of different binary-gate-type neurons.
Networks with AND gate neurons consistently exhibit the slowest increase in accuracy.
Among those with OR and XOR gates, the lead in rate of increase shifts depending on the number of neurons involved.
However, a consistent finding is that the maximum achievable accuracy, when equal numbers of neurons are engaged, is comparable across all tested network configurations with different types of gate neurons.
This holds true even when the input lattice size, $n_{\text{in}}$, is set to the maximal tested value~of~160.
The variation in the average maximum values obtained from the simulations does not exceed 0.2\%, which is significantly less than the variance due to the use of different randomization masks.
This finding underscores the fact that structural nonlinearity, emerging due to the network’s layout, plays a pivotal role in the network's functionality.

\section{Discussion and conclusions}

In our study, we rigorously explored the functionality of a binarized neuromorphic network based on a lattice of exciton-polariton dyads, pairs of optically excited polariton condensates within a microcavity that function as binary logic gate neurons.
These dyads are controlled by optically altering the potential landscape to toggle between different ON and OFF states, effectively implementing binary gate functions such as NAND, NOR, and XNOR.
We employed numerical simulations to investigate the impact of these binary neuron types on the network's ability to classify images from the MNIST dataset, serving as a benchmark to assess the effectiveness of different neuronal configurations. Our analyses focused on the relationship between network structure, the nonlinearity of neuron responses, and overall system performance, assessing how changes in input signal density and neuron configuration affected classification accuracy.

Our research has revealed the intricate balance between structural and inherent nonlinearity within polariton-based binarized neuromorphic networks.
The results emphasize that the pivotal role in the network's functionality is played by structural nonlinearity, emerging due to the network's layout.
This nonlinearity ensures that our networks can handle complex computational tasks effectively, even when the individual neurons are not inherently nonlinear.
While nonlinear neurons are a compelling physical concept, the nonlinear response of individual neurons in our proposed architecture proves non-essential for high performance.
Such behavior is primarily pronounced when the number of neurons involved in computations is low, a condition under which any neural network operates suboptimally.
This finding challenges the traditional emphasis on the necessity of inherent nonlinearity within neurons for effective neural computation.
Instead, our results suggest that the configuration and interaction of elements within the network can simulate the benefits of nonlinearity, effectively bypassing the need for complex neuron functions.

The implications of these findings are profound, suggesting that future research and development in neuromorphic computing could focus more on network design rather than solely on the properties of individual neurons.
This shift could lead to more efficient designs that are simpler to manufacture and operate, particularly beneficial for scalable deployment in various applications.
Earlier assessments of potential energy reductions in polariton-based neuromorphic networks, made in~\cite{NanoLett213715,PhysRevApplied16024045}, predominantly emphasized the need for significant nonlinearity within individual neurons to ensure effective operation.
Our results challenge this approach by demonstrating that individual neuron nonlinearity is not essential, which could lead to further substantial reductions in energy consumption estimates.

\begin{acknowledgments}
The support of Saint-Petersburg State University research project 122040800257-5 and
the  state assignment in the field of scientific activity of the Ministry of Science and Higher Education of the Russian Federation (theme FZUN-2024-0019, state assignment of the VlSU)  are acknowledged. 
\end{acknowledgments}

\appendix

\section*{Appendix: Modelling of polariton dyads}

For simulating ensembles of polariton dyads, we use the generalized Gross-Pitaevskii equation for the polariton wave function $\Psi(t,\mathbf{r})$:
\begin{equation}
\label{EqGPE}
i\hbar \partial _t \Psi (t,\mathbf{r}) = \left[-\frac{\hbar ^2 }{2 m^*} \nabla^2 + U(t,\mathbf{r}) - \frac{i \hbar \gamma}{2} \right] \Psi (t,\mathbf{r}),
\end{equation}
where $m^*$ is the effective polariton mass, $\gamma$ is the polariton decay rate.
$U(t,\mathbf{r})$ is an effective potential for polaritons that can be written in the following form: 
\begin{equation}
\label{EqPotential}
U(t,\mathbf{r}) = V(\mathbf{r}) + \alpha |\Psi (t,\mathbf{r})|^2 + U_{\text{d}}(t,\mathbf{r}) +   G P_{\text{s}} (\mathbf{r}).
\end{equation}
The first term in the right hand side of Eq.~\eqref{EqPotential} characterizes the stationary potential for isolating dyads.
For simulations, we take it in the form analogous to one in Ref.~\cite{sedov2024polarlattneur} as $V (\mathbf{r}) = F(\mathbf{r}, V_0, u, a) $, where $F$ is a function of a spatial coordinate ($\mathbf{r}$), with $V_0$, $u$ and $a$ being the height of the potential, the  width of the walls, the width of the gap in the wall, responsible for entrance of the input signal, respectively.
The second term characterizes polariton-polariton interactions with the interaction constant~$\alpha$.
The third term is responsible for the complex effective potential arising from the non-resonant optical pumping of the condensates in the dyads.
It can be written as
\begin{equation}
\label{EqPotentiald} U_{\text{d}}(t,\mathbf{r}) = \frac{(g + i R)P (\mathbf{r})}{2(\gamma_{\text{R}} + R |\Psi (t,\mathbf{r})|^2)} ,
\end{equation}
where $P(\mathbf{r})$ is the non-resonant optical pumping intensity for exciting polariton condensates in dyads,
$g$ is the constant of interaction of polaritons with excitons in the bright reservoir,
$R$ is the stimulated scattering rate from the reservoir to the condensate,
$\gamma _{\text{R}}$ is the decay rate of reservoir excitons.

The last term in Eq.~\eqref{EqPotential} is responsible for the potential, arising from the reservoir of dark excitons, induced by the signal beams of intensity~$P_{\text{s}}(\mathbf{r})$. 
$G$~is~a constant, that characterizes repulsion from the dark reservoir barrier.

For simulating the ensemble of polariton dyads depicted in Fig.~\ref{FIG_Gates}, we take the optical pumps as ensembles of Gaussian beams in the following form:
\begin{subequations}
\begin{multline}
 P (\mathbf{r}) \propto  \sum _{j,k} \exp \left\{ -[x + (j+0.5) d]^2 / 2 w_{x}^2 \right. \\
 \left.+ [y + (k + 0.5) d]^2 / 2 w_{y}^2 \right\}, 
\end{multline}
\begin{multline}
P_{\text{s}} (\mathbf{r}) \propto \sum _{j,k} \exp \left[ -(x+jd)^2 / 2 w_{\text{s},x}^2 \right. \\
 \left.+ (y +k d)^2 / 2 w_{\text{s},y}^2 \right],
\end{multline}
\end{subequations}
where $j,k \in \mathbb{Z}$ numerate positions of the pump spots,
$w_{x,y}$ and $w_{\text{s},x,y}$ are widths of the spots.

For simulation, we take the following values of the parameters, matching those in~\cite{PhysRevLett124207402,sedov2024polarlattneur}.
The effective polariton mass is $m^* = 0.49 \, \text{meV ps}^2 \mu\text{m}^{-2}$, 
the decay rates are ${\gamma = 1/6 \, \text{ps}^{-1}}$ and ${\gamma _{\text{R}} = 0.05 \, \text{ps}^{-1}}$,
the interaction constants are ${\alpha = 2.4 \, \mu \text{eV} \, \mu \text{m}^2}$, ${g = 4 \alpha}$,
the scattering rate is ${R = 7 \alpha}$. 
The parameters for the optical pumps have been chosen as follows.
The lattice period is $d=10 \,\mu\text{m}$, the pump spot widths are $w_{x,y} = 1.8\, \mu \text{m}$ and $w_{\text{s}, x,y} = 2\, \mu \text{m}$ for the NOR-gate neurons,
$d=10 \,\mu\text{m}$, $w_{x,y} = 1.8\, \mu \text{m}$ and $w_{\text{s}, x,y} = 1.65\, \mu \text{m}$ for the NAND-gate neurons,
$d=15 \,\mu\text{m}$, $w_{x,y} = 1.7\, \mu \text{m}$ and  ${w_{\text{s}, x} = 0.6w_{\text{s}, x} = 5.1\, \mu \text{m}}$ for the XNOR-gate neurons.

\bibliographystyle{apsrev4-1}
\bibliography{polaritonBrainGatesBibl}

\begin{thebibliography}{44}%
\makeatletter
\providecommand \@ifxundefined [1]{%
 \@ifx{#1\undefined}
}%
\providecommand \@ifnum [1]{%
 \ifnum #1\expandafter \@firstoftwo
 \else \expandafter \@secondoftwo
 \fi
}%
\providecommand \@ifx [1]{%
 \ifx #1\expandafter \@firstoftwo
 \else \expandafter \@secondoftwo
 \fi
}%
\providecommand \natexlab [1]{#1}%
\providecommand \enquote  [1]{``#1''}%
\providecommand \bibnamefont  [1]{#1}%
\providecommand \bibfnamefont [1]{#1}%
\providecommand \citenamefont [1]{#1}%
\providecommand \href@noop [0]{\@secondoftwo}%
\providecommand \href [0]{\begingroup \@sanitize@url \@href}%
\providecommand \@href[1]{\@@startlink{#1}\@@href}%
\providecommand \@@href[1]{\endgroup#1\@@endlink}%
\providecommand \@sanitize@url [0]{\catcode `\\12\catcode `\$12\catcode
  `\&12\catcode `\#12\catcode `\^12\catcode `\_12\catcode `\%12\relax}%
\providecommand \@@startlink[1]{}%
\providecommand \@@endlink[0]{}%
\providecommand \url  [0]{\begingroup\@sanitize@url \@url }%
\providecommand \@url [1]{\endgroup\@href {#1}{\urlprefix }}%
\providecommand \urlprefix  [0]{URL }%
\providecommand \Eprint [0]{\href }%
\providecommand \doibase [0]{http://dx.doi.org/}%
\providecommand \selectlanguage [0]{\@gobble}%
\providecommand \bibinfo  [0]{\@secondoftwo}%
\providecommand \bibfield  [0]{\@secondoftwo}%
\providecommand \translation [1]{[#1]}%
\providecommand \BibitemOpen [0]{}%
\providecommand \bibitemStop [0]{}%
\providecommand \bibitemNoStop [0]{.\EOS\space}%
\providecommand \EOS [0]{\spacefactor3000\relax}%
\providecommand \BibitemShut  [1]{\csname bibitem#1\endcsname}%
\let\auto@bib@innerbib\@empty
\bibitem [{\citenamefont {Ozmen}\ \emph {et~al.}(2021)\citenamefont {Ozmen},
  \citenamefont {Ozmen},\ and\ \citenamefont {Ko{\c{c}}}}]{Ozmen2021}%
  \BibitemOpen
  \bibfield  {author} {\bibinfo {author} {\bibfnamefont {M.~M.}\ \bibnamefont
  {Ozmen}}, \bibinfo {author} {\bibfnamefont {A.}~\bibnamefont {Ozmen}}, \ and\
  \bibinfo {author} {\bibfnamefont {{\c{C}}.~K.}\ \bibnamefont {Ko{\c{c}}}},\
  }\enquote {\bibinfo {title} {Artificial intelligence for next-generation
  medical robotics},}\ in\ \href {\doibase 10.1007/978-3-030-49100-0_3} {\emph
  {\bibinfo {booktitle} {Digital Surgery}}},\ \bibinfo {editor} {edited by\
  \bibinfo {editor} {\bibfnamefont {S.}~\bibnamefont {Atallah}}}\ (\bibinfo
  {publisher} {Springer International Publishing},\ \bibinfo {address} {Cham},\
  \bibinfo {year} {2021})\ pp.\ \bibinfo {pages} {25--36}\BibitemShut {NoStop}%
\bibitem [{\citenamefont {Bekbolatova}\ \emph {et~al.}(2024)\citenamefont
  {Bekbolatova}, \citenamefont {Mayer}, \citenamefont {Ong},\ and\
  \citenamefont {Toma}}]{healthcare12020125}%
  \BibitemOpen
  \bibfield  {author} {\bibinfo {author} {\bibfnamefont {M.}~\bibnamefont
  {Bekbolatova}}, \bibinfo {author} {\bibfnamefont {J.}~\bibnamefont {Mayer}},
  \bibinfo {author} {\bibfnamefont {C.~W.}\ \bibnamefont {Ong}}, \ and\
  \bibinfo {author} {\bibfnamefont {M.}~\bibnamefont {Toma}},\ }\href {\doibase
  10.3390/healthcare12020125} {\bibfield  {journal} {\bibinfo  {journal}
  {Healthcare}\ }\textbf {\bibinfo {volume} {12}} (\bibinfo {year} {2024}),\
  10.3390/healthcare12020125}\BibitemShut {NoStop}%
\bibitem [{\citenamefont {Zhu}\ \emph {et~al.}(2020)\citenamefont {Zhu},
  \citenamefont {Zhang}, \citenamefont {Yang},\ and\ \citenamefont
  {Huang}}]{APRes7011312}%
  \BibitemOpen
  \bibfield  {author} {\bibinfo {author} {\bibfnamefont {J.}~\bibnamefont
  {Zhu}}, \bibinfo {author} {\bibfnamefont {T.}~\bibnamefont {Zhang}}, \bibinfo
  {author} {\bibfnamefont {Y.}~\bibnamefont {Yang}}, \ and\ \bibinfo {author}
  {\bibfnamefont {R.}~\bibnamefont {Huang}},\ }\href {\doibase
  10.1063/1.5118217} {\bibfield  {journal} {\bibinfo  {journal} {Applied
  Physics Reviews}\ }\textbf {\bibinfo {volume} {7}},\ \bibinfo {pages}
  {011312} (\bibinfo {year} {2020})}\BibitemShut {NoStop}%
\bibitem [{\citenamefont {Merolla}\ \emph {et~al.}(2014)\citenamefont
  {Merolla}, \citenamefont {Arthur}, \citenamefont {Alvarez-Icaza},
  \citenamefont {Cassidy}, \citenamefont {Sawada}, \citenamefont {Akopyan},
  \citenamefont {Jackson}, \citenamefont {Imam}, \citenamefont {Guo},
  \citenamefont {Nakamura}, \citenamefont {Brezzo}, \citenamefont {Vo},
  \citenamefont {Esser}, \citenamefont {Appuswamy}, \citenamefont {Taba},
  \citenamefont {Amir}, \citenamefont {Flickner}, \citenamefont {Risk},
  \citenamefont {Manohar},\ and\ \citenamefont {Modha}}]{Science345668}%
  \BibitemOpen
  \bibfield  {author} {\bibinfo {author} {\bibfnamefont {P.~A.}\ \bibnamefont
  {Merolla}}, \bibinfo {author} {\bibfnamefont {J.~V.}\ \bibnamefont {Arthur}},
  \bibinfo {author} {\bibfnamefont {R.}~\bibnamefont {Alvarez-Icaza}}, \bibinfo
  {author} {\bibfnamefont {A.~S.}\ \bibnamefont {Cassidy}}, \bibinfo {author}
  {\bibfnamefont {J.}~\bibnamefont {Sawada}}, \bibinfo {author} {\bibfnamefont
  {F.}~\bibnamefont {Akopyan}}, \bibinfo {author} {\bibfnamefont {B.~L.}\
  \bibnamefont {Jackson}}, \bibinfo {author} {\bibfnamefont {N.}~\bibnamefont
  {Imam}}, \bibinfo {author} {\bibfnamefont {C.}~\bibnamefont {Guo}}, \bibinfo
  {author} {\bibfnamefont {Y.}~\bibnamefont {Nakamura}}, \bibinfo {author}
  {\bibfnamefont {B.}~\bibnamefont {Brezzo}}, \bibinfo {author} {\bibfnamefont
  {I.}~\bibnamefont {Vo}}, \bibinfo {author} {\bibfnamefont {S.~K.}\
  \bibnamefont {Esser}}, \bibinfo {author} {\bibfnamefont {R.}~\bibnamefont
  {Appuswamy}}, \bibinfo {author} {\bibfnamefont {B.}~\bibnamefont {Taba}},
  \bibinfo {author} {\bibfnamefont {A.}~\bibnamefont {Amir}}, \bibinfo {author}
  {\bibfnamefont {M.~D.}\ \bibnamefont {Flickner}}, \bibinfo {author}
  {\bibfnamefont {W.~P.}\ \bibnamefont {Risk}}, \bibinfo {author}
  {\bibfnamefont {R.}~\bibnamefont {Manohar}}, \ and\ \bibinfo {author}
  {\bibfnamefont {D.~S.}\ \bibnamefont {Modha}},\ }\href {\doibase
  10.1126/science.1254642} {\bibfield  {journal} {\bibinfo  {journal}
  {Science}\ }\textbf {\bibinfo {volume} {345}},\ \bibinfo {pages} {668}
  (\bibinfo {year} {2014})}\BibitemShut {NoStop}%
\bibitem [{\citenamefont {Ballarini}\ \emph {et~al.}(2020)\citenamefont
  {Ballarini}, \citenamefont {Gianfrate}, \citenamefont {Panico}, \citenamefont
  {Opala}, \citenamefont {Ghosh}, \citenamefont {Dominici}, \citenamefont
  {Ardizzone}, \citenamefont {De~Giorgi}, \citenamefont {Lerario},
  \citenamefont {Gigli}, \citenamefont {Liew}, \citenamefont {Matuszewski},\
  and\ \citenamefont {Sanvitto}}]{Nanolett203506}%
  \BibitemOpen
  \bibfield  {author} {\bibinfo {author} {\bibfnamefont {D.}~\bibnamefont
  {Ballarini}}, \bibinfo {author} {\bibfnamefont {A.}~\bibnamefont
  {Gianfrate}}, \bibinfo {author} {\bibfnamefont {R.}~\bibnamefont {Panico}},
  \bibinfo {author} {\bibfnamefont {A.}~\bibnamefont {Opala}}, \bibinfo
  {author} {\bibfnamefont {S.}~\bibnamefont {Ghosh}}, \bibinfo {author}
  {\bibfnamefont {L.}~\bibnamefont {Dominici}}, \bibinfo {author}
  {\bibfnamefont {V.}~\bibnamefont {Ardizzone}}, \bibinfo {author}
  {\bibfnamefont {M.}~\bibnamefont {De~Giorgi}}, \bibinfo {author}
  {\bibfnamefont {G.}~\bibnamefont {Lerario}}, \bibinfo {author} {\bibfnamefont
  {G.}~\bibnamefont {Gigli}}, \bibinfo {author} {\bibfnamefont {T.~C.~H.}\
  \bibnamefont {Liew}}, \bibinfo {author} {\bibfnamefont {M.}~\bibnamefont
  {Matuszewski}}, \ and\ \bibinfo {author} {\bibfnamefont {D.}~\bibnamefont
  {Sanvitto}},\ }\href {\doibase 10.1021/acs.nanolett.0c00435} {\bibfield
  {journal} {\bibinfo  {journal} {Nano Letters}\ }\textbf {\bibinfo {volume}
  {20}},\ \bibinfo {pages} {3506} (\bibinfo {year} {2020})}\BibitemShut
  {NoStop}%
\bibitem [{\citenamefont {Opala}\ \emph {et~al.}(2022)\citenamefont {Opala},
  \citenamefont {Panico}, \citenamefont {Ardizzone}, \citenamefont {Pietka},
  \citenamefont {Szczytko}, \citenamefont {Sanvitto}, \citenamefont
  {Matuszewski},\ and\ \citenamefont {Ballarini}}]{PhysRevApplied18024028}%
  \BibitemOpen
  \bibfield  {author} {\bibinfo {author} {\bibfnamefont {A.}~\bibnamefont
  {Opala}}, \bibinfo {author} {\bibfnamefont {R.}~\bibnamefont {Panico}},
  \bibinfo {author} {\bibfnamefont {V.}~\bibnamefont {Ardizzone}}, \bibinfo
  {author} {\bibfnamefont {B.}~\bibnamefont {Pietka}}, \bibinfo {author}
  {\bibfnamefont {J.}~\bibnamefont {Szczytko}}, \bibinfo {author}
  {\bibfnamefont {D.}~\bibnamefont {Sanvitto}}, \bibinfo {author}
  {\bibfnamefont {M.}~\bibnamefont {Matuszewski}}, \ and\ \bibinfo {author}
  {\bibfnamefont {D.}~\bibnamefont {Ballarini}},\ }\href {\doibase
  10.1103/PhysRevApplied.18.024028} {\bibfield  {journal} {\bibinfo  {journal}
  {Phys. Rev. Appl.}\ }\textbf {\bibinfo {volume} {18}},\ \bibinfo {pages}
  {024028} (\bibinfo {year} {2022})}\BibitemShut {NoStop}%
\bibitem [{\citenamefont {Tait}\ \emph {et~al.}(2017)\citenamefont {Tait},
  \citenamefont {de~Lima}, \citenamefont {Zhou}, \citenamefont {Wu},
  \citenamefont {Nahmias}, \citenamefont {Shastri},\ and\ \citenamefont
  {Prucnal}}]{SciRep77430}%
  \BibitemOpen
  \bibfield  {author} {\bibinfo {author} {\bibfnamefont {A.~N.}\ \bibnamefont
  {Tait}}, \bibinfo {author} {\bibfnamefont {T.~F.}\ \bibnamefont {de~Lima}},
  \bibinfo {author} {\bibfnamefont {E.}~\bibnamefont {Zhou}}, \bibinfo {author}
  {\bibfnamefont {A.~X.}\ \bibnamefont {Wu}}, \bibinfo {author} {\bibfnamefont
  {M.~A.}\ \bibnamefont {Nahmias}}, \bibinfo {author} {\bibfnamefont {B.~J.}\
  \bibnamefont {Shastri}}, \ and\ \bibinfo {author} {\bibfnamefont {P.~R.}\
  \bibnamefont {Prucnal}},\ }\href {\doibase 10.1038/s41598-017-07754-z}
  {\bibfield  {journal} {\bibinfo  {journal} {Scientific Reports}\ }\textbf
  {\bibinfo {volume} {7}},\ \bibinfo {pages} {7430} (\bibinfo {year}
  {2017})}\BibitemShut {NoStop}%
\bibitem [{\citenamefont {Ancora}\ \emph {et~al.}(2024)\citenamefont {Ancora},
  \citenamefont {Negri}, \citenamefont {Gianfrate}, \citenamefont
  {Trypogeorgos}, \citenamefont {Dominici}, \citenamefont {Sanvitto},
  \citenamefont {Ricci-Tersenghi},\ and\ \citenamefont
  {Leuzzi}}]{PhysRevApplied21064027}%
  \BibitemOpen
  \bibfield  {author} {\bibinfo {author} {\bibfnamefont {D.}~\bibnamefont
  {Ancora}}, \bibinfo {author} {\bibfnamefont {M.}~\bibnamefont {Negri}},
  \bibinfo {author} {\bibfnamefont {A.}~\bibnamefont {Gianfrate}}, \bibinfo
  {author} {\bibfnamefont {D.}~\bibnamefont {Trypogeorgos}}, \bibinfo {author}
  {\bibfnamefont {L.}~\bibnamefont {Dominici}}, \bibinfo {author}
  {\bibfnamefont {D.}~\bibnamefont {Sanvitto}}, \bibinfo {author}
  {\bibfnamefont {F.}~\bibnamefont {Ricci-Tersenghi}}, \ and\ \bibinfo {author}
  {\bibfnamefont {L.}~\bibnamefont {Leuzzi}},\ }\href {\doibase
  10.1103/PhysRevApplied.21.064027} {\bibfield  {journal} {\bibinfo  {journal}
  {Phys. Rev. Appl.}\ }\textbf {\bibinfo {volume} {21}},\ \bibinfo {pages}
  {064027} (\bibinfo {year} {2024})}\BibitemShut {NoStop}%
\bibitem [{\citenamefont {Mirek}\ \emph {et~al.}(2021)\citenamefont {Mirek},
  \citenamefont {Opala}, \citenamefont {Comaron}, \citenamefont {Furman},
  \citenamefont {Kr{\'o}l}, \citenamefont {Tyszka}, \citenamefont
  {Seredy{\'n}ski}, \citenamefont {Ballarini}, \citenamefont {Sanvitto},
  \citenamefont {Liew}, \citenamefont {Pacuski}, \citenamefont
  {Suffczy{\'n}ski}, \citenamefont {Szczytko}, \citenamefont {Matuszewski},\
  and\ \citenamefont {Pietka}}]{NanoLett213715}%
  \BibitemOpen
  \bibfield  {author} {\bibinfo {author} {\bibfnamefont {R.}~\bibnamefont
  {Mirek}}, \bibinfo {author} {\bibfnamefont {A.}~\bibnamefont {Opala}},
  \bibinfo {author} {\bibfnamefont {P.}~\bibnamefont {Comaron}}, \bibinfo
  {author} {\bibfnamefont {M.}~\bibnamefont {Furman}}, \bibinfo {author}
  {\bibfnamefont {M.}~\bibnamefont {Kr{\'o}l}}, \bibinfo {author}
  {\bibfnamefont {K.}~\bibnamefont {Tyszka}}, \bibinfo {author} {\bibfnamefont
  {B.}~\bibnamefont {Seredy{\'n}ski}}, \bibinfo {author} {\bibfnamefont
  {D.}~\bibnamefont {Ballarini}}, \bibinfo {author} {\bibfnamefont
  {D.}~\bibnamefont {Sanvitto}}, \bibinfo {author} {\bibfnamefont {T.~C.~H.}\
  \bibnamefont {Liew}}, \bibinfo {author} {\bibfnamefont {W.}~\bibnamefont
  {Pacuski}}, \bibinfo {author} {\bibfnamefont {J.}~\bibnamefont
  {Suffczy{\'n}ski}}, \bibinfo {author} {\bibfnamefont {J.}~\bibnamefont
  {Szczytko}}, \bibinfo {author} {\bibfnamefont {M.}~\bibnamefont
  {Matuszewski}}, \ and\ \bibinfo {author} {\bibfnamefont {B.}~\bibnamefont
  {Pietka}},\ }\href {\doibase 10.1021/acs.nanolett.0c04696} {\bibfield
  {journal} {\bibinfo  {journal} {Nano Letters}\ }\textbf {\bibinfo {volume}
  {21}},\ \bibinfo {pages} {3715} (\bibinfo {year} {2021})}\BibitemShut
  {NoStop}%
\bibitem [{\citenamefont {Matuszewski}\ \emph {et~al.}(2021)\citenamefont
  {Matuszewski}, \citenamefont {Opala}, \citenamefont {Mirek}, \citenamefont
  {Furman}, \citenamefont {Kr\'ol}, \citenamefont {Tyszka}, \citenamefont
  {Liew}, \citenamefont {Ballarini}, \citenamefont {Sanvitto}, \citenamefont
  {Szczytko},\ and\ \citenamefont {Pietka}}]{PhysRevApplied16024045}%
  \BibitemOpen
  \bibfield  {author} {\bibinfo {author} {\bibfnamefont {M.}~\bibnamefont
  {Matuszewski}}, \bibinfo {author} {\bibfnamefont {A.}~\bibnamefont {Opala}},
  \bibinfo {author} {\bibfnamefont {R.}~\bibnamefont {Mirek}}, \bibinfo
  {author} {\bibfnamefont {M.}~\bibnamefont {Furman}}, \bibinfo {author}
  {\bibfnamefont {M.}~\bibnamefont {Kr\'ol}}, \bibinfo {author} {\bibfnamefont
  {K.}~\bibnamefont {Tyszka}}, \bibinfo {author} {\bibfnamefont
  {T.}~\bibnamefont {Liew}}, \bibinfo {author} {\bibfnamefont {D.}~\bibnamefont
  {Ballarini}}, \bibinfo {author} {\bibfnamefont {D.}~\bibnamefont {Sanvitto}},
  \bibinfo {author} {\bibfnamefont {J.}~\bibnamefont {Szczytko}}, \ and\
  \bibinfo {author} {\bibfnamefont {B.}~\bibnamefont {Pietka}},\ }\href
  {\doibase 10.1103/PhysRevApplied.16.024045} {\bibfield  {journal} {\bibinfo
  {journal} {Phys. Rev. Appl.}\ }\textbf {\bibinfo {volume} {16}},\ \bibinfo
  {pages} {024045} (\bibinfo {year} {2021})}\BibitemShut {NoStop}%
\bibitem [{\citenamefont {Hopfield}(1958)}]{PhysRev1121555}%
  \BibitemOpen
  \bibfield  {author} {\bibinfo {author} {\bibfnamefont {J.~J.}\ \bibnamefont
  {Hopfield}},\ }\href {\doibase 10.1103/PhysRev.112.1555} {\bibfield
  {journal} {\bibinfo  {journal} {Phys. Rev.}\ }\textbf {\bibinfo {volume}
  {112}},\ \bibinfo {pages} {1555} (\bibinfo {year} {1958})}\BibitemShut
  {NoStop}%
\bibitem [{\citenamefont {Hopfield}(1969)}]{PhysRev182945}%
  \BibitemOpen
  \bibfield  {author} {\bibinfo {author} {\bibfnamefont {J.~J.}\ \bibnamefont
  {Hopfield}},\ }\href {\doibase 10.1103/PhysRev.182.945} {\bibfield  {journal}
  {\bibinfo  {journal} {Phys. Rev.}\ }\textbf {\bibinfo {volume} {182}},\
  \bibinfo {pages} {945} (\bibinfo {year} {1969})}\BibitemShut {NoStop}%
\bibitem [{\citenamefont {Hopfield}(1982)}]{Hopfield1982aa}%
  \BibitemOpen
  \bibfield  {author} {\bibinfo {author} {\bibfnamefont {J.~J.}\ \bibnamefont
  {Hopfield}},\ }\href {\doibase 10.1073/pnas.79.8.2554} {\bibfield  {journal}
  {\bibinfo  {journal} {Proc Natl Acad Sci U S A}\ }\textbf {\bibinfo {volume}
  {79}},\ \bibinfo {pages} {2554} (\bibinfo {year} {1982})}\BibitemShut
  {NoStop}%
\bibitem [{\citenamefont {Hopfield}\ and\ \citenamefont
  {Tank}(1986)}]{science3755256}%
  \BibitemOpen
  \bibfield  {author} {\bibinfo {author} {\bibfnamefont {J.~J.}\ \bibnamefont
  {Hopfield}}\ and\ \bibinfo {author} {\bibfnamefont {D.~W.}\ \bibnamefont
  {Tank}},\ }\href {\doibase 10.1126/science.3755256} {\bibfield  {journal}
  {\bibinfo  {journal} {Science}\ }\textbf {\bibinfo {volume} {233}},\ \bibinfo
  {pages} {625} (\bibinfo {year} {1986})}\BibitemShut {NoStop}%
\bibitem [{\citenamefont {Simons}\ and\ \citenamefont
  {Lee}(2019)}]{electronics8060661}%
  \BibitemOpen
  \bibfield  {author} {\bibinfo {author} {\bibfnamefont {T.}~\bibnamefont
  {Simons}}\ and\ \bibinfo {author} {\bibfnamefont {D.-J.}\ \bibnamefont
  {Lee}},\ }\href {\doibase 10.3390/electronics8060661} {\bibfield  {journal}
  {\bibinfo  {journal} {Electronics}\ }\textbf {\bibinfo {volume} {8}}
  (\bibinfo {year} {2019}),\ 10.3390/electronics8060661}\BibitemShut {NoStop}%
\bibitem [{\citenamefont {Kim}\ \emph {et~al.}(2019)\citenamefont {Kim},
  \citenamefont {Choi}, \citenamefont {Yoon}, \citenamefont {Lee},
  \citenamefont {Kim}, \citenamefont {Kang},\ and\ \citenamefont
  {Choi}}]{SciRep911705}%
  \BibitemOpen
  \bibfield  {author} {\bibinfo {author} {\bibfnamefont {S.}~\bibnamefont
  {Kim}}, \bibinfo {author} {\bibfnamefont {B.}~\bibnamefont {Choi}}, \bibinfo
  {author} {\bibfnamefont {J.}~\bibnamefont {Yoon}}, \bibinfo {author}
  {\bibfnamefont {Y.}~\bibnamefont {Lee}}, \bibinfo {author} {\bibfnamefont
  {H.-D.}\ \bibnamefont {Kim}}, \bibinfo {author} {\bibfnamefont {M.-H.}\
  \bibnamefont {Kang}}, \ and\ \bibinfo {author} {\bibfnamefont {S.-J.}\
  \bibnamefont {Choi}},\ }\href {\doibase 10.1038/s41598-019-48048-w}
  {\bibfield  {journal} {\bibinfo  {journal} {Scientific Reports}\ }\textbf
  {\bibinfo {volume} {9}},\ \bibinfo {pages} {11705} (\bibinfo {year}
  {2019})}\BibitemShut {NoStop}%
\bibitem [{\citenamefont {Qin}\ \emph {et~al.}(2020)\citenamefont {Qin},
  \citenamefont {Gong}, \citenamefont {Liu}, \citenamefont {Bai}, \citenamefont
  {Song},\ and\ \citenamefont {Sebe}}]{QIN2020107281}%
  \BibitemOpen
  \bibfield  {author} {\bibinfo {author} {\bibfnamefont {H.}~\bibnamefont
  {Qin}}, \bibinfo {author} {\bibfnamefont {R.}~\bibnamefont {Gong}}, \bibinfo
  {author} {\bibfnamefont {X.}~\bibnamefont {Liu}}, \bibinfo {author}
  {\bibfnamefont {X.}~\bibnamefont {Bai}}, \bibinfo {author} {\bibfnamefont
  {J.}~\bibnamefont {Song}}, \ and\ \bibinfo {author} {\bibfnamefont
  {N.}~\bibnamefont {Sebe}},\ }\href {\doibase
  https://doi.org/10.1016/j.patcog.2020.107281} {\bibfield  {journal} {\bibinfo
   {journal} {Pattern Recognition}\ }\textbf {\bibinfo {volume} {105}},\
  \bibinfo {pages} {107281} (\bibinfo {year} {2020})}\BibitemShut {NoStop}%
\bibitem [{\citenamefont {Chi}\ and\ \citenamefont
  {Jiang}(2021)}]{Chi2021Logic}%
  \BibitemOpen
  \bibfield  {author} {\bibinfo {author} {\bibfnamefont {C.-C.}\ \bibnamefont
  {Chi}}\ and\ \bibinfo {author} {\bibfnamefont {J.~H.}\ \bibnamefont
  {Jiang}},\ }\href {\doibase 10.1109/TCAD.2021.3078606} {\bibfield  {journal}
  {\bibinfo  {journal} {IEEE Transactions on Computer-Aided Design of
  Integrated Circuits and Systems}\ }\textbf {\bibinfo {volume} {41}},\
  \bibinfo {pages} {993} (\bibinfo {year} {2021})}\BibitemShut {NoStop}%
\bibitem [{\citenamefont {Liang}\ \emph {et~al.}(2018)\citenamefont {Liang},
  \citenamefont {Yin}, \citenamefont {Liu}, \citenamefont {Luk},\ and\
  \citenamefont {Wei}}]{Liang2018FPBNN}%
  \BibitemOpen
  \bibfield  {author} {\bibinfo {author} {\bibfnamefont {S.}~\bibnamefont
  {Liang}}, \bibinfo {author} {\bibfnamefont {S.}~\bibnamefont {Yin}}, \bibinfo
  {author} {\bibfnamefont {L.}~\bibnamefont {Liu}}, \bibinfo {author}
  {\bibfnamefont {W.}~\bibnamefont {Luk}}, \ and\ \bibinfo {author}
  {\bibfnamefont {S.}~\bibnamefont {Wei}},\ }\href {\doibase
  10.1016/j.neucom.2017.09.046} {\bibfield  {journal} {\bibinfo  {journal}
  {Neurocomputing}\ }\textbf {\bibinfo {volume} {275}},\ \bibinfo {pages}
  {1072} (\bibinfo {year} {2018})}\BibitemShut {NoStop}%
\bibitem [{\citenamefont {Dubey}\ \emph {et~al.}(2022)\citenamefont {Dubey},
  \citenamefont {Singh},\ and\ \citenamefont {Chaudhuri}}]{DUBEY202292}%
  \BibitemOpen
  \bibfield  {author} {\bibinfo {author} {\bibfnamefont {S.~R.}\ \bibnamefont
  {Dubey}}, \bibinfo {author} {\bibfnamefont {S.~K.}\ \bibnamefont {Singh}}, \
  and\ \bibinfo {author} {\bibfnamefont {B.~B.}\ \bibnamefont {Chaudhuri}},\
  }\href {\doibase https://doi.org/10.1016/j.neucom.2022.06.111} {\bibfield
  {journal} {\bibinfo  {journal} {Neurocomputing}\ }\textbf {\bibinfo {volume}
  {503}},\ \bibinfo {pages} {92} (\bibinfo {year} {2022})}\BibitemShut
  {NoStop}%
\bibitem [{\citenamefont {Wanjura}\ and\ \citenamefont
  {Marquardt}(2024)}]{NatPhysWanjura2024}%
  \BibitemOpen
  \bibfield  {author} {\bibinfo {author} {\bibfnamefont {C.~C.}\ \bibnamefont
  {Wanjura}}\ and\ \bibinfo {author} {\bibfnamefont {F.}~\bibnamefont
  {Marquardt}},\ }\href {\doibase 10.1038/s41567-024-02534-9} {\bibfield
  {journal} {\bibinfo  {journal} {Nature Physics}\ } (\bibinfo {year} {2024}),\
  10.1038/s41567-024-02534-9}\BibitemShut {NoStop}%
\bibitem [{\citenamefont {Eliezer}\ \emph {et~al.}(2023)\citenamefont
  {Eliezer}, \citenamefont {Rührmair}, \citenamefont {Wisiol}, \citenamefont
  {Bittner},\ and\ \citenamefont {Cao}}]{PNAS120e2305027120}%
  \BibitemOpen
  \bibfield  {author} {\bibinfo {author} {\bibfnamefont {Y.}~\bibnamefont
  {Eliezer}}, \bibinfo {author} {\bibfnamefont {U.}~\bibnamefont {Rührmair}},
  \bibinfo {author} {\bibfnamefont {N.}~\bibnamefont {Wisiol}}, \bibinfo
  {author} {\bibfnamefont {S.}~\bibnamefont {Bittner}}, \ and\ \bibinfo
  {author} {\bibfnamefont {H.}~\bibnamefont {Cao}},\ }\href {\doibase
  10.1073/pnas.2305027120} {\bibfield  {journal} {\bibinfo  {journal}
  {Proceedings of the National Academy of Sciences}\ }\textbf {\bibinfo
  {volume} {120}},\ \bibinfo {pages} {e2305027120} (\bibinfo {year}
  {2023})}\BibitemShut {NoStop}%
\bibitem [{\citenamefont {Xia}\ \emph {et~al.}(2023)\citenamefont {Xia},
  \citenamefont {Kim}, \citenamefont {Eliezer}, \citenamefont {Shaughnessy},
  \citenamefont {Gigan},\ and\ \citenamefont
  {Cao}}]{xia2023deeplearningpassiveoptical}%
  \BibitemOpen
  \bibfield  {author} {\bibinfo {author} {\bibfnamefont {F.}~\bibnamefont
  {Xia}}, \bibinfo {author} {\bibfnamefont {K.}~\bibnamefont {Kim}}, \bibinfo
  {author} {\bibfnamefont {Y.}~\bibnamefont {Eliezer}}, \bibinfo {author}
  {\bibfnamefont {L.}~\bibnamefont {Shaughnessy}}, \bibinfo {author}
  {\bibfnamefont {S.}~\bibnamefont {Gigan}}, \ and\ \bibinfo {author}
  {\bibfnamefont {H.}~\bibnamefont {Cao}},\ }\href
  {https://arxiv.org/abs/2307.08558} {\enquote {\bibinfo {title} {Deep learning
  with passive optical nonlinear mapping},}\ } (\bibinfo {year} {2023}),\
  \Eprint {http://arxiv.org/abs/2307.08558} {arXiv:2307.08558 [physics.optics]}
  \BibitemShut {NoStop}%
\bibitem [{\citenamefont {Yildirim}\ \emph {et~al.}(2024)\citenamefont
  {Yildirim}, \citenamefont {Dinc}, \citenamefont {Oguz}, \citenamefont
  {Psaltis},\ and\ \citenamefont
  {Moser}}]{yildirim2024nonlinearprocessinglinearoptics}%
  \BibitemOpen
  \bibfield  {author} {\bibinfo {author} {\bibfnamefont {M.}~\bibnamefont
  {Yildirim}}, \bibinfo {author} {\bibfnamefont {N.~U.}\ \bibnamefont {Dinc}},
  \bibinfo {author} {\bibfnamefont {I.}~\bibnamefont {Oguz}}, \bibinfo {author}
  {\bibfnamefont {D.}~\bibnamefont {Psaltis}}, \ and\ \bibinfo {author}
  {\bibfnamefont {C.}~\bibnamefont {Moser}},\ }\href
  {https://arxiv.org/abs/2307.08533} {\enquote {\bibinfo {title} {Nonlinear
  processing with linear optics},}\ } (\bibinfo {year} {2024}),\ \Eprint
  {http://arxiv.org/abs/2307.08533} {arXiv:2307.08533 [physics.optics]}
  \BibitemShut {NoStop}%
\bibitem [{\citenamefont {McMahon}(2024)}]{NatPhysMcMahon2024}%
  \BibitemOpen
  \bibfield  {author} {\bibinfo {author} {\bibfnamefont {P.~L.}\ \bibnamefont
  {McMahon}},\ }\href {\doibase 10.1038/s41567-024-02531-y} {\bibfield
  {journal} {\bibinfo  {journal} {Nature Physics}\ } (\bibinfo {year} {2024}),\
  10.1038/s41567-024-02531-y}\BibitemShut {NoStop}%
\bibitem [{\citenamefont {Farmakidis}\ \emph {et~al.}(2024)\citenamefont
  {Farmakidis}, \citenamefont {Dong},\ and\ \citenamefont
  {Bhaskaran}}]{NatRevElEng1358}%
  \BibitemOpen
  \bibfield  {author} {\bibinfo {author} {\bibfnamefont {N.}~\bibnamefont
  {Farmakidis}}, \bibinfo {author} {\bibfnamefont {B.}~\bibnamefont {Dong}}, \
  and\ \bibinfo {author} {\bibfnamefont {H.}~\bibnamefont {Bhaskaran}},\ }\href
  {\doibase 10.1038/s44287-024-00050-9} {\bibfield  {journal} {\bibinfo
  {journal} {Nature Reviews Electrical Engineering}\ }\textbf {\bibinfo
  {volume} {1}},\ \bibinfo {pages} {358} (\bibinfo {year} {2024})}\BibitemShut
  {NoStop}%
\bibitem [{\citenamefont {Sedov}\ and\ \citenamefont
  {Kavokin}(2024)}]{sedov2024polarlattneur}%
  \BibitemOpen
  \bibfield  {author} {\bibinfo {author} {\bibfnamefont {E.}~\bibnamefont
  {Sedov}}\ and\ \bibinfo {author} {\bibfnamefont {A.}~\bibnamefont
  {Kavokin}},\ }\href {https://arxiv.org/abs/2401.07232} {\enquote {\bibinfo
  {title} {Polariton lattices as binarized neuromorphic networks},}\ }
  (\bibinfo {year} {2024}),\ \Eprint {http://arxiv.org/abs/2401.07232}
  {arXiv:2401.07232 [cond-mat.dis-nn]} \BibitemShut {NoStop}%
\bibitem [{\citenamefont {Alyatkin}\ \emph {et~al.}(2020)\citenamefont
  {Alyatkin}, \citenamefont {T\"opfer}, \citenamefont {Askitopoulos},
  \citenamefont {Sigurdsson},\ and\ \citenamefont
  {Lagoudakis}}]{PhysRevLett124207402}%
  \BibitemOpen
  \bibfield  {author} {\bibinfo {author} {\bibfnamefont {S.}~\bibnamefont
  {Alyatkin}}, \bibinfo {author} {\bibfnamefont {J.~D.}\ \bibnamefont
  {T\"opfer}}, \bibinfo {author} {\bibfnamefont {A.}~\bibnamefont
  {Askitopoulos}}, \bibinfo {author} {\bibfnamefont {H.}~\bibnamefont
  {Sigurdsson}}, \ and\ \bibinfo {author} {\bibfnamefont {P.~G.}\ \bibnamefont
  {Lagoudakis}},\ }\href {\doibase 10.1103/PhysRevLett.124.207402} {\bibfield
  {journal} {\bibinfo  {journal} {Phys. Rev. Lett.}\ }\textbf {\bibinfo
  {volume} {124}},\ \bibinfo {pages} {207402} (\bibinfo {year}
  {2020})}\BibitemShut {NoStop}%
\bibitem [{\citenamefont {Kasprzak}\ \emph {et~al.}(2006)\citenamefont
  {Kasprzak}, \citenamefont {Richard}, \citenamefont {Kundermann},
  \citenamefont {Baas}, \citenamefont {Jeambrun}, \citenamefont {Keeling},
  \citenamefont {Marchetti}, \citenamefont {Szyma{\'n}ska}, \citenamefont
  {Andr{\'e}}, \citenamefont {Staehli}, \citenamefont {Savona}, \citenamefont
  {Littlewood}, \citenamefont {Deveaud},\ and\ \citenamefont
  {Dang}}]{Nature443409}%
  \BibitemOpen
  \bibfield  {author} {\bibinfo {author} {\bibfnamefont {J.}~\bibnamefont
  {Kasprzak}}, \bibinfo {author} {\bibfnamefont {M.}~\bibnamefont {Richard}},
  \bibinfo {author} {\bibfnamefont {S.}~\bibnamefont {Kundermann}}, \bibinfo
  {author} {\bibfnamefont {A.}~\bibnamefont {Baas}}, \bibinfo {author}
  {\bibfnamefont {P.}~\bibnamefont {Jeambrun}}, \bibinfo {author}
  {\bibfnamefont {J.~M.~J.}\ \bibnamefont {Keeling}}, \bibinfo {author}
  {\bibfnamefont {F.~M.}\ \bibnamefont {Marchetti}}, \bibinfo {author}
  {\bibfnamefont {M.~H.}\ \bibnamefont {Szyma{\'n}ska}}, \bibinfo {author}
  {\bibfnamefont {R.}~\bibnamefont {Andr{\'e}}}, \bibinfo {author}
  {\bibfnamefont {J.~L.}\ \bibnamefont {Staehli}}, \bibinfo {author}
  {\bibfnamefont {V.}~\bibnamefont {Savona}}, \bibinfo {author} {\bibfnamefont
  {P.~B.}\ \bibnamefont {Littlewood}}, \bibinfo {author} {\bibfnamefont
  {B.}~\bibnamefont {Deveaud}}, \ and\ \bibinfo {author} {\bibfnamefont
  {L.~S.}\ \bibnamefont {Dang}},\ }\href {\doibase 10.1038/nature05131}
  {\bibfield  {journal} {\bibinfo  {journal} {Nature}\ }\textbf {\bibinfo
  {volume} {443}},\ \bibinfo {pages} {409} (\bibinfo {year}
  {2006})}\BibitemShut {NoStop}%
\bibitem [{\citenamefont {Sedov}\ \emph {et~al.}(2020)\citenamefont {Sedov},
  \citenamefont {Lukoshkin}, \citenamefont {Kalevich}, \citenamefont
  {Hatzopoulos}, \citenamefont {Savvidis},\ and\ \citenamefont
  {Kavokin}}]{ACSPhot71163}%
  \BibitemOpen
  \bibfield  {author} {\bibinfo {author} {\bibfnamefont {E.}~\bibnamefont
  {Sedov}}, \bibinfo {author} {\bibfnamefont {V.}~\bibnamefont {Lukoshkin}},
  \bibinfo {author} {\bibfnamefont {V.}~\bibnamefont {Kalevich}}, \bibinfo
  {author} {\bibfnamefont {Z.}~\bibnamefont {Hatzopoulos}}, \bibinfo {author}
  {\bibfnamefont {P.}~\bibnamefont {Savvidis}}, \ and\ \bibinfo {author}
  {\bibfnamefont {A.}~\bibnamefont {Kavokin}},\ }\href {\doibase
  10.1021/acsphotonics.9b01779} {\bibfield  {journal} {\bibinfo  {journal} {ACS
  Photonics}\ }\textbf {\bibinfo {volume} {7}},\ \bibinfo {pages} {1163}
  (\bibinfo {year} {2020})}\BibitemShut {NoStop}%
\bibitem [{\citenamefont {T\"{o}pfer}\ \emph {et~al.}(2021)\citenamefont
  {T\"{o}pfer}, \citenamefont {Chatzopoulos}, \citenamefont {Sigurdsson},
  \citenamefont {Cookson}, \citenamefont {Rubo},\ and\ \citenamefont
  {Lagoudakis}}]{Optica8106}%
  \BibitemOpen
  \bibfield  {author} {\bibinfo {author} {\bibfnamefont {J.~D.}\ \bibnamefont
  {T\"{o}pfer}}, \bibinfo {author} {\bibfnamefont {I.}~\bibnamefont
  {Chatzopoulos}}, \bibinfo {author} {\bibfnamefont {H.}~\bibnamefont
  {Sigurdsson}}, \bibinfo {author} {\bibfnamefont {T.}~\bibnamefont {Cookson}},
  \bibinfo {author} {\bibfnamefont {Y.~G.}\ \bibnamefont {Rubo}}, \ and\
  \bibinfo {author} {\bibfnamefont {P.~G.}\ \bibnamefont {Lagoudakis}},\ }\href
  {\doibase 10.1364/OPTICA.409976} {\bibfield  {journal} {\bibinfo  {journal}
  {Optica}\ }\textbf {\bibinfo {volume} {8}},\ \bibinfo {pages} {106} (\bibinfo
  {year} {2021})}\BibitemShut {NoStop}%
\bibitem [{\citenamefont {Renucci}\ \emph {et~al.}(2005)\citenamefont
  {Renucci}, \citenamefont {Amand}, \citenamefont {Marie}, \citenamefont
  {Senellart}, \citenamefont {Bloch}, \citenamefont {Sermage},\ and\
  \citenamefont {Kavokin}}]{PhysRevB72075317}%
  \BibitemOpen
  \bibfield  {author} {\bibinfo {author} {\bibfnamefont {P.}~\bibnamefont
  {Renucci}}, \bibinfo {author} {\bibfnamefont {T.}~\bibnamefont {Amand}},
  \bibinfo {author} {\bibfnamefont {X.}~\bibnamefont {Marie}}, \bibinfo
  {author} {\bibfnamefont {P.}~\bibnamefont {Senellart}}, \bibinfo {author}
  {\bibfnamefont {J.}~\bibnamefont {Bloch}}, \bibinfo {author} {\bibfnamefont
  {B.}~\bibnamefont {Sermage}}, \ and\ \bibinfo {author} {\bibfnamefont
  {K.~V.}\ \bibnamefont {Kavokin}},\ }\href {\doibase
  10.1103/PhysRevB.72.075317} {\bibfield  {journal} {\bibinfo  {journal} {Phys.
  Rev. B}\ }\textbf {\bibinfo {volume} {72}},\ \bibinfo {pages} {075317}
  (\bibinfo {year} {2005})}\BibitemShut {NoStop}%
\bibitem [{\citenamefont {Takemura}\ \emph {et~al.}(2014)\citenamefont
  {Takemura}, \citenamefont {Trebaol}, \citenamefont {Wouters}, \citenamefont
  {Portella-Oberli},\ and\ \citenamefont {Deveaud}}]{PhysRevB90195307}%
  \BibitemOpen
  \bibfield  {author} {\bibinfo {author} {\bibfnamefont {N.}~\bibnamefont
  {Takemura}}, \bibinfo {author} {\bibfnamefont {S.}~\bibnamefont {Trebaol}},
  \bibinfo {author} {\bibfnamefont {M.}~\bibnamefont {Wouters}}, \bibinfo
  {author} {\bibfnamefont {M.~T.}\ \bibnamefont {Portella-Oberli}}, \ and\
  \bibinfo {author} {\bibfnamefont {B.}~\bibnamefont {Deveaud}},\ }\href
  {\doibase 10.1103/PhysRevB.90.195307} {\bibfield  {journal} {\bibinfo
  {journal} {Phys. Rev. B}\ }\textbf {\bibinfo {volume} {90}},\ \bibinfo
  {pages} {195307} (\bibinfo {year} {2014})}\BibitemShut {NoStop}%
\bibitem [{\citenamefont {del Valle-Inclan~Redondo}\ \emph
  {et~al.}(2018)\citenamefont {del Valle-Inclan~Redondo}, \citenamefont
  {Ohadi}, \citenamefont {Rubo}, \citenamefont {Beer}, \citenamefont {Ramsay},
  \citenamefont {Tsintzos}, \citenamefont {Hatzopoulos}, \citenamefont
  {Savvidis},\ and\ \citenamefont {Baumberg}}]{NJPhys20075008}%
  \BibitemOpen
  \bibfield  {author} {\bibinfo {author} {\bibfnamefont {Y.}~\bibnamefont {del
  Valle-Inclan~Redondo}}, \bibinfo {author} {\bibfnamefont {H.}~\bibnamefont
  {Ohadi}}, \bibinfo {author} {\bibfnamefont {Y.~G.}\ \bibnamefont {Rubo}},
  \bibinfo {author} {\bibfnamefont {O.}~\bibnamefont {Beer}}, \bibinfo {author}
  {\bibfnamefont {A.~J.}\ \bibnamefont {Ramsay}}, \bibinfo {author}
  {\bibfnamefont {S.~I.}\ \bibnamefont {Tsintzos}}, \bibinfo {author}
  {\bibfnamefont {Z.}~\bibnamefont {Hatzopoulos}}, \bibinfo {author}
  {\bibfnamefont {P.~G.}\ \bibnamefont {Savvidis}}, \ and\ \bibinfo {author}
  {\bibfnamefont {J.~J.}\ \bibnamefont {Baumberg}},\ }\href {\doibase
  10.1088/1367-2630/aad377} {\bibfield  {journal} {\bibinfo  {journal} {New
  Journal of Physics}\ }\textbf {\bibinfo {volume} {20}},\ \bibinfo {pages}
  {075008} (\bibinfo {year} {2018})}\BibitemShut {NoStop}%
\bibitem [{\citenamefont {Balas}\ \emph {et~al.}(2022)\citenamefont {Balas},
  \citenamefont {Sedov}, \citenamefont {Paschos}, \citenamefont {Hatzopoulos},
  \citenamefont {Ohadi}, \citenamefont {Kavokin},\ and\ \citenamefont
  {Savvidis}}]{PhysRevLett128117401}%
  \BibitemOpen
  \bibfield  {author} {\bibinfo {author} {\bibfnamefont {Y.~C.}\ \bibnamefont
  {Balas}}, \bibinfo {author} {\bibfnamefont {E.~S.}\ \bibnamefont {Sedov}},
  \bibinfo {author} {\bibfnamefont {G.~G.}\ \bibnamefont {Paschos}}, \bibinfo
  {author} {\bibfnamefont {Z.}~\bibnamefont {Hatzopoulos}}, \bibinfo {author}
  {\bibfnamefont {H.}~\bibnamefont {Ohadi}}, \bibinfo {author} {\bibfnamefont
  {A.~V.}\ \bibnamefont {Kavokin}}, \ and\ \bibinfo {author} {\bibfnamefont
  {P.~G.}\ \bibnamefont {Savvidis}},\ }\href {\doibase
  10.1103/PhysRevLett.128.117401} {\bibfield  {journal} {\bibinfo  {journal}
  {Phys. Rev. Lett.}\ }\textbf {\bibinfo {volume} {128}},\ \bibinfo {pages}
  {117401} (\bibinfo {year} {2022})}\BibitemShut {NoStop}%
\bibitem [{\citenamefont {Schmidt}\ \emph {et~al.}(2019)\citenamefont
  {Schmidt}, \citenamefont {Berger}, \citenamefont {Kahlert}, \citenamefont
  {Bayer}, \citenamefont {Schneider}, \citenamefont {H\"ofling}, \citenamefont
  {Sedov}, \citenamefont {Kavokin},\ and\ \citenamefont
  {A\ss{}mann}}]{PhysRevLett122047403}%
  \BibitemOpen
  \bibfield  {author} {\bibinfo {author} {\bibfnamefont {D.}~\bibnamefont
  {Schmidt}}, \bibinfo {author} {\bibfnamefont {B.}~\bibnamefont {Berger}},
  \bibinfo {author} {\bibfnamefont {M.}~\bibnamefont {Kahlert}}, \bibinfo
  {author} {\bibfnamefont {M.}~\bibnamefont {Bayer}}, \bibinfo {author}
  {\bibfnamefont {C.}~\bibnamefont {Schneider}}, \bibinfo {author}
  {\bibfnamefont {S.}~\bibnamefont {H\"ofling}}, \bibinfo {author}
  {\bibfnamefont {E.~S.}\ \bibnamefont {Sedov}}, \bibinfo {author}
  {\bibfnamefont {A.~V.}\ \bibnamefont {Kavokin}}, \ and\ \bibinfo {author}
  {\bibfnamefont {M.}~\bibnamefont {A\ss{}mann}},\ }\href {\doibase
  10.1103/PhysRevLett.122.047403} {\bibfield  {journal} {\bibinfo  {journal}
  {Phys. Rev. Lett.}\ }\textbf {\bibinfo {volume} {122}},\ \bibinfo {pages}
  {047403} (\bibinfo {year} {2019})}\BibitemShut {NoStop}%
\bibitem [{\citenamefont {Rozas}\ \emph {et~al.}(2023)\citenamefont {Rozas},
  \citenamefont {Sedov}, \citenamefont {Brune}, \citenamefont {H\"ofling},
  \citenamefont {Kavokin},\ and\ \citenamefont
  {A\ss{}mann}}]{PhysRevB108165411}%
  \BibitemOpen
  \bibfield  {author} {\bibinfo {author} {\bibfnamefont {E.}~\bibnamefont
  {Rozas}}, \bibinfo {author} {\bibfnamefont {E.}~\bibnamefont {Sedov}},
  \bibinfo {author} {\bibfnamefont {Y.}~\bibnamefont {Brune}}, \bibinfo
  {author} {\bibfnamefont {S.}~\bibnamefont {H\"ofling}}, \bibinfo {author}
  {\bibfnamefont {A.}~\bibnamefont {Kavokin}}, \ and\ \bibinfo {author}
  {\bibfnamefont {M.}~\bibnamefont {A\ss{}mann}},\ }\href {\doibase
  10.1103/PhysRevB.108.165411} {\bibfield  {journal} {\bibinfo  {journal}
  {Phys. Rev. B}\ }\textbf {\bibinfo {volume} {108}},\ \bibinfo {pages}
  {165411} (\bibinfo {year} {2023})}\BibitemShut {NoStop}%
\bibitem [{\citenamefont {Askitopoulos}\ \emph {et~al.}(2015)\citenamefont
  {Askitopoulos}, \citenamefont {Liew}, \citenamefont {Ohadi}, \citenamefont
  {Hatzopoulos}, \citenamefont {Savvidis},\ and\ \citenamefont
  {Lagoudakis}}]{PhysRevB92035305}%
  \BibitemOpen
  \bibfield  {author} {\bibinfo {author} {\bibfnamefont {A.}~\bibnamefont
  {Askitopoulos}}, \bibinfo {author} {\bibfnamefont {T.~C.~H.}\ \bibnamefont
  {Liew}}, \bibinfo {author} {\bibfnamefont {H.}~\bibnamefont {Ohadi}},
  \bibinfo {author} {\bibfnamefont {Z.}~\bibnamefont {Hatzopoulos}}, \bibinfo
  {author} {\bibfnamefont {P.~G.}\ \bibnamefont {Savvidis}}, \ and\ \bibinfo
  {author} {\bibfnamefont {P.~G.}\ \bibnamefont {Lagoudakis}},\ }\href
  {\doibase 10.1103/PhysRevB.92.035305} {\bibfield  {journal} {\bibinfo
  {journal} {Phys. Rev. B}\ }\textbf {\bibinfo {volume} {92}},\ \bibinfo
  {pages} {035305} (\bibinfo {year} {2015})}\BibitemShut {NoStop}%
\bibitem [{\citenamefont {Askitopoulos}\ \emph {et~al.}(2018)\citenamefont
  {Askitopoulos}, \citenamefont {Nalitov}, \citenamefont {Sedov}, \citenamefont
  {Pickup}, \citenamefont {Cherotchenko}, \citenamefont {Hatzopoulos},
  \citenamefont {Savvidis}, \citenamefont {Kavokin},\ and\ \citenamefont
  {Lagoudakis}}]{PhysRevB97235303}%
  \BibitemOpen
  \bibfield  {author} {\bibinfo {author} {\bibfnamefont {A.}~\bibnamefont
  {Askitopoulos}}, \bibinfo {author} {\bibfnamefont {A.~V.}\ \bibnamefont
  {Nalitov}}, \bibinfo {author} {\bibfnamefont {E.~S.}\ \bibnamefont {Sedov}},
  \bibinfo {author} {\bibfnamefont {L.}~\bibnamefont {Pickup}}, \bibinfo
  {author} {\bibfnamefont {E.~D.}\ \bibnamefont {Cherotchenko}}, \bibinfo
  {author} {\bibfnamefont {Z.}~\bibnamefont {Hatzopoulos}}, \bibinfo {author}
  {\bibfnamefont {P.~G.}\ \bibnamefont {Savvidis}}, \bibinfo {author}
  {\bibfnamefont {A.~V.}\ \bibnamefont {Kavokin}}, \ and\ \bibinfo {author}
  {\bibfnamefont {P.~G.}\ \bibnamefont {Lagoudakis}},\ }\href {\doibase
  10.1103/PhysRevB.97.235303} {\bibfield  {journal} {\bibinfo  {journal} {Phys.
  Rev. B}\ }\textbf {\bibinfo {volume} {97}},\ \bibinfo {pages} {235303}
  (\bibinfo {year} {2018})}\BibitemShut {NoStop}%
\bibitem [{\citenamefont {Aladinskaia}\ \emph {et~al.}(2023)\citenamefont
  {Aladinskaia}, \citenamefont {Cherbunin}, \citenamefont {Sedov},
  \citenamefont {Liubomirov}, \citenamefont {Kavokin}, \citenamefont
  {Khramtsov}, \citenamefont {Petrov}, \citenamefont {Savvidis},\ and\
  \citenamefont {Kavokin}}]{PhysRevB107045302}%
  \BibitemOpen
  \bibfield  {author} {\bibinfo {author} {\bibfnamefont {E.}~\bibnamefont
  {Aladinskaia}}, \bibinfo {author} {\bibfnamefont {R.}~\bibnamefont
  {Cherbunin}}, \bibinfo {author} {\bibfnamefont {E.}~\bibnamefont {Sedov}},
  \bibinfo {author} {\bibfnamefont {A.}~\bibnamefont {Liubomirov}}, \bibinfo
  {author} {\bibfnamefont {K.}~\bibnamefont {Kavokin}}, \bibinfo {author}
  {\bibfnamefont {E.}~\bibnamefont {Khramtsov}}, \bibinfo {author}
  {\bibfnamefont {M.}~\bibnamefont {Petrov}}, \bibinfo {author} {\bibfnamefont
  {P.~G.}\ \bibnamefont {Savvidis}}, \ and\ \bibinfo {author} {\bibfnamefont
  {A.}~\bibnamefont {Kavokin}},\ }\href {\doibase 10.1103/PhysRevB.107.045302}
  {\bibfield  {journal} {\bibinfo  {journal} {Phys. Rev. B}\ }\textbf {\bibinfo
  {volume} {107}},\ \bibinfo {pages} {045302} (\bibinfo {year}
  {2023})}\BibitemShut {NoStop}%
\bibitem [{\citenamefont {Galbiati}\ \emph {et~al.}(2012)\citenamefont
  {Galbiati}, \citenamefont {Ferrier}, \citenamefont {Solnyshkov},
  \citenamefont {Tanese}, \citenamefont {Wertz}, \citenamefont {Amo},
  \citenamefont {Abbarchi}, \citenamefont {Senellart}, \citenamefont {Sagnes},
  \citenamefont {Lema\^{\i}tre}, \citenamefont {Galopin}, \citenamefont
  {Malpuech},\ and\ \citenamefont {Bloch}}]{PhysRevLett108126403}%
  \BibitemOpen
  \bibfield  {author} {\bibinfo {author} {\bibfnamefont {M.}~\bibnamefont
  {Galbiati}}, \bibinfo {author} {\bibfnamefont {L.}~\bibnamefont {Ferrier}},
  \bibinfo {author} {\bibfnamefont {D.~D.}\ \bibnamefont {Solnyshkov}},
  \bibinfo {author} {\bibfnamefont {D.}~\bibnamefont {Tanese}}, \bibinfo
  {author} {\bibfnamefont {E.}~\bibnamefont {Wertz}}, \bibinfo {author}
  {\bibfnamefont {A.}~\bibnamefont {Amo}}, \bibinfo {author} {\bibfnamefont
  {M.}~\bibnamefont {Abbarchi}}, \bibinfo {author} {\bibfnamefont
  {P.}~\bibnamefont {Senellart}}, \bibinfo {author} {\bibfnamefont
  {I.}~\bibnamefont {Sagnes}}, \bibinfo {author} {\bibfnamefont
  {A.}~\bibnamefont {Lema\^{\i}tre}}, \bibinfo {author} {\bibfnamefont
  {E.}~\bibnamefont {Galopin}}, \bibinfo {author} {\bibfnamefont
  {G.}~\bibnamefont {Malpuech}}, \ and\ \bibinfo {author} {\bibfnamefont
  {J.}~\bibnamefont {Bloch}},\ }\href {\doibase 10.1103/PhysRevLett.108.126403}
  {\bibfield  {journal} {\bibinfo  {journal} {Phys. Rev. Lett.}\ }\textbf
  {\bibinfo {volume} {108}},\ \bibinfo {pages} {126403} (\bibinfo {year}
  {2012})}\BibitemShut {NoStop}%
\bibitem [{\citenamefont {Baboux}\ \emph {et~al.}(2016)\citenamefont {Baboux},
  \citenamefont {Ge}, \citenamefont {Jacqmin}, \citenamefont {Biondi},
  \citenamefont {Galopin}, \citenamefont {Lema\^{\i}tre}, \citenamefont
  {Le~Gratiet}, \citenamefont {Sagnes}, \citenamefont {Schmidt}, \citenamefont
  {T\"ureci}, \citenamefont {Amo},\ and\ \citenamefont
  {Bloch}}]{PhysRevLett116066402}%
  \BibitemOpen
  \bibfield  {author} {\bibinfo {author} {\bibfnamefont {F.}~\bibnamefont
  {Baboux}}, \bibinfo {author} {\bibfnamefont {L.}~\bibnamefont {Ge}}, \bibinfo
  {author} {\bibfnamefont {T.}~\bibnamefont {Jacqmin}}, \bibinfo {author}
  {\bibfnamefont {M.}~\bibnamefont {Biondi}}, \bibinfo {author} {\bibfnamefont
  {E.}~\bibnamefont {Galopin}}, \bibinfo {author} {\bibfnamefont
  {A.}~\bibnamefont {Lema\^{\i}tre}}, \bibinfo {author} {\bibfnamefont
  {L.}~\bibnamefont {Le~Gratiet}}, \bibinfo {author} {\bibfnamefont
  {I.}~\bibnamefont {Sagnes}}, \bibinfo {author} {\bibfnamefont
  {S.}~\bibnamefont {Schmidt}}, \bibinfo {author} {\bibfnamefont {H.~E.}\
  \bibnamefont {T\"ureci}}, \bibinfo {author} {\bibfnamefont {A.}~\bibnamefont
  {Amo}}, \ and\ \bibinfo {author} {\bibfnamefont {J.}~\bibnamefont {Bloch}},\
  }\href {\doibase 10.1103/PhysRevLett.116.066402} {\bibfield  {journal}
  {\bibinfo  {journal} {Phys. Rev. Lett.}\ }\textbf {\bibinfo {volume} {116}},\
  \bibinfo {pages} {066402} (\bibinfo {year} {2016})}\BibitemShut {NoStop}%
\bibitem [{\citenamefont {Kalinin}\ and\ \citenamefont
  {Berloff}(2020)}]{AdvQT31900065}%
  \BibitemOpen
  \bibfield  {author} {\bibinfo {author} {\bibfnamefont {K.~P.}\ \bibnamefont
  {Kalinin}}\ and\ \bibinfo {author} {\bibfnamefont {N.~G.}\ \bibnamefont
  {Berloff}},\ }\href {\doibase https://doi.org/10.1002/qute.201900065}
  {\bibfield  {journal} {\bibinfo  {journal} {Advanced Quantum Technologies}\
  }\textbf {\bibinfo {volume} {3}},\ \bibinfo {pages} {1900065} (\bibinfo
  {year} {2020})}\BibitemShut {NoStop}%
\bibitem [{\citenamefont {Schneider}\ \emph {et~al.}(2016)\citenamefont
  {Schneider}, \citenamefont {Winkler}, \citenamefont {Fraser}, \citenamefont
  {Kamp}, \citenamefont {Yamamoto}, \citenamefont {Ostrovskaya},\ and\
  \citenamefont {Höfling}}]{RepProgrPhys80016503}%
  \BibitemOpen
  \bibfield  {author} {\bibinfo {author} {\bibfnamefont {C.}~\bibnamefont
  {Schneider}}, \bibinfo {author} {\bibfnamefont {K.}~\bibnamefont {Winkler}},
  \bibinfo {author} {\bibfnamefont {M.~D.}\ \bibnamefont {Fraser}}, \bibinfo
  {author} {\bibfnamefont {M.}~\bibnamefont {Kamp}}, \bibinfo {author}
  {\bibfnamefont {Y.}~\bibnamefont {Yamamoto}}, \bibinfo {author}
  {\bibfnamefont {E.~A.}\ \bibnamefont {Ostrovskaya}}, \ and\ \bibinfo {author}
  {\bibfnamefont {S.}~\bibnamefont {Höfling}},\ }\href {\doibase
  10.1088/0034-4885/80/1/016503} {\bibfield  {journal} {\bibinfo  {journal}
  {Reports on Progress in Physics}\ }\textbf {\bibinfo {volume} {80}},\
  \bibinfo {pages} {016503} (\bibinfo {year} {2016})}\BibitemShut {NoStop}%
\end{thebibliography}%

\end{document}